\documentclass[sn-apa]{sn-jnl}

\usepackage[normalem]{ulem}
\useunder{\uline}{\ul}{}
\usepackage{booktabs}
\usepackage{multirow}
\usepackage{tablefootnote}
\usepackage{array}
\usepackage{ragged2e}
\usepackage{lscape}
\usepackage{tikz}
\usepackage{program}
\usepackage{rotating}
\usepackage{makecell, caption}
\usepackage{url}
\usepackage{multirow}
\usepackage{arydshln}
\usepackage{xcolor}

\newcommand*{\CirclOK}[1][1]{%
  \begin{tikzpicture}[every path/.style={thick,fill=lightgray},scale=#1]
  \tiny
  \draw[fill=lime] (0,0) circle (0.15) node {};
  \end{tikzpicture}%
}

\newcommand*{\RectTEC}[1][1]{%
  \begin{tikzpicture}[every path/.style={thick,fill=lightgray},scale=#1]
  \tiny
  \draw[fill=cyan] (0,0) rectangle (0.3,0.3) node[pos=.5] {T};
  \end{tikzpicture}%
}

\newcommand*{\CircleHigh}[1][1]{%
  \begin{tikzpicture}[every path/.style={thick,fill=lightgray},scale=#1]
  \tiny
  \draw[fill=orange] (0,0) circle (0.08) node {};
  \end{tikzpicture}%
}

\newcommand*{\CircleLikert}[1][1]{%
  \begin{tikzpicture}[every path/.style={thick,fill=lightgray},scale=#1]
  \tiny
  \draw[fill=white, very thin] (0,0) circle (0.08) node {};
  \end{tikzpicture}%
}

\newcommand*{\CircleLikertFake}[1][1]{%
  \begin{tikzpicture}[every path/.style={fill=white},scale=#1]
  \tiny
  \draw[white] (0,0) circle (0.08) node {};
  \end{tikzpicture}%
}

\jyear{2021}%

\theoremstyle{thmstyleone}%
\theoremstyle{thmstyletwo}%
\theoremstyle{thmstylethree}%
\begin{document}

\title[Use case cards]{Use case cards: a use case reporting framework inspired by the European AI Act}



\author*[1]{\fnm{Isabelle} \sur{Hupont}}\email{isabelle.hupont-torres@ec.europa.eu}

\author[1,2]{\fnm{David} \sur{Fernández-Llorca}}\email{david.fernandez-llorca@ec.europa.eu}

\author[3]{\fnm{Sandra} \sur{Baldassarri}}\email{sandra@unizar.es}

\author[1]{\fnm{Emilia} \sur{Gómez}}\email{emilia.gomez-gutierrez@ec.europa.eu}

\affil*[1]{\orgdiv{Joint Research Centre}, \orgname{European Commission}, \city{Seville}, \country{Spain}}

\affil[2]{\orgdiv{Computer Engineering Department}, \orgname{University of Alcalá}, \city{Alcalá de Henares}, \country{Spain}}

\affil[3]{\orgdiv{Instituto Universitario de Investigación en Ingenería de Aragón}, \orgname{University of Zaragoza}, \city{Zaragoza},  \country{Spain}}


\abstract{
Despite recent efforts by the Artificial Intelligence (AI) community to move towards standardised procedures for documenting models, methods, systems or datasets, there is currently no methodology focused on use cases aligned with the risk-based approach of the European AI Act (AI Act). In this paper, we propose a new framework for the documentation of use cases, that we call \textit{use case cards}, based on the use case modelling included in the Unified Markup Language (UML) standard. Unlike other documentation methodologies, we focus on the intended purpose and operational use of an AI system. It consists of two main parts. Firstly, a UML-based template, tailored to allow implicitly assessing the risk level of the AI system and defining relevant requirements. Secondly, a supporting UML diagram designed to provide information about the system-user  interactions and relationships. The proposed framework is the result of a co-design process involving a relevant team of EU policy experts and scientists.
We have validated our proposal with 11 experts with different backgrounds and a reasonable knowledge of the AI Act as a prerequisite. We provide the 5 \textit{use case cards} used in the co-design and validation process. 
\textit{Use case cards} allows framing and contextualising use cases in an effective way, and we hope this methodology can be a useful tool for policy makers and providers for documenting use cases, assessing the risk level, adapting the different requirements and building a catalogue of existing usages of AI.
}




\maketitle


\section{Introduction}

Nowadays, Artificial Intelligence (AI) is living a groundbreaking moment from many perspectives, including the technological, societal and legal ones. On the one hand, more and more powerful and technologically mature AI systems are being used by the wide public on a daily basis, including recommender systems, decision-support systems, content generation systems, person identification and object recognition systems, and conversational systems. On the other hand, policy makers are starting to put in place legal grounds aiming at regulating the trustworthy use of AI.

With this exponential trend in the daily use of AI, there is need to put in place robust mechanisms to foster a better understanding of AI systems by all impacted stakeholders --experts and non-experts-- in order to help ensuring their trustworthy, safe and fair use. Indeed, several studies have acknowledged that the issue of how to communicate about the functioning and potential limits of increasingly complex AI systems remains an open challenge~\citep{laato2022explain}.

In particular, transparency in the form of well-structured documentation practices is considered a key step towards trustworthy AI~\citep{HLEG}. Some methodologies for AI documentation have emerged and been rapidly adopted in the recent years. Nevertheless, their target audience is typically AI technical practitioners (e.g. AI developers, designers, data scientists) leaving aside other important personas such as policy makers or citizens~\citep{hupont2022documenting}. Moreover, the focus is mainly put on technical characteristics (e.g. performance, representativity) of the data used for training~\citep{gebru2018datasheets} and/or general-purpose AI models~\citep{mitchell2019model}. 
When it comes to document more specific \textit{use cases} of AI systems, i.e. a real-world deployment of an AI system in a concrete operational environment and for a particular purpose, documentation is generally limited to a brief textual description without a standardised format~\citep{louradour2021policy}. 

Nowadays, voluntary AI documentation practices are in the process of becoming legal requirements in some countries. The recent European Commission's proposal for the Regulation of Artificial Intelligence, the AI Act~\citep{AIact}, aims at regulating software systems that are developed with AI techniques such as machine 
learning. Interestingly, the legal text does not mandate concrete technical solutions to be adopted; instead, it focuses on the \textit{intended purpose} of an AI system which determines its risk profile and, consequently, a set of legal requirements that must be met. Thus, the AI Act's approach further reinforces the need to properly cover the documentation of AI use cases, which are directly related to the intended purpose of an AI system.

The technique of \textit{use case modelling} has been used for decades in classic software development~\citep{cockburn2001writing}. The so-modelled use cases provide insights into how different actors interact with a software system, the user interface design and the main system's components. 
It allows developers to identify the system's boundaries and required functionalities, ensuring that all stakeholders are satisfied and have a shared understanding of the system's expected behaviour~\citep{fantechi2003applications}. The use case modelling technique therefore serves as a common mean of communication between stakeholders, including developers, designers, testers, business analysts, clients and end users, allowing for effective collaboration and reducing misunderstandings with respect to functional requirements. 

Building upon some preliminary work focusing on the affective computing domain \citep{hupont2022documenting2}, this study revisits classic software use case modelling methodologies, more specifically the widely-used Unified Markup Language (UML) specification~\citep{UML251}, to propose a standardised template-based approach for AI use case documentation: the \textit{use case card}. To ensure that  \textit{use case cards} cover all the information needs required for the assessment of use cases through the lenses of the European AI Act, the methodology has been developed following a co-design process involving European Commission's AI policy experts, AI scientific officers and an external UML and User Experience (UX) expert. Several examples of \textit{use case cards} are then validated in a user study to check for adequateness, completeness and usability. The \textit{use case card} template and all implemented examples are publicly available at the GitLab repository {\url{https://gitlab.com/humaint-ec_public/use-case-cards}}.


The remainder of the paper is as follows. Section~\ref{sec:background} reviews the central role of use cases within the AI Act, identifies the needs in terms of information elements for their documentation, and reflects on how current AI documentation methodologies fail to cover these needs. Section~\ref{sec:methodology} presents the \textit{use case card} documentation methodology and details how to fill it. Section~\ref{sec:user_study} elaborates on the co-design process and validation of \textit{use case cards} with key stakeholders. Finally, Section~\ref{sec:conclusions} concludes the work.

\section{Background}
\label{sec:background}

\subsection{The central role of use cases in the AI policy context}
\label{subsec:aia_intended}

An AI model is a mathematical algorithm designed to perform a computational task. It is generally trained using large datasets and machine learning techniques, from which it learns patterns and relationships to make predictions or generate outputs when presented with new input. Popular examples of AI models include object detectors, language/image generation models or content search algorithms. AI models are typically created in a controlled environment, such as a research lab. 
At this stage an AI model is in most cases generic, meaning that a very same model can be used for many different purposes. For instance, an object detector can be embedded in car's software system to recognise vehicles, road signs and pedestrians~\citep{gupta2021deep}, or be used for automatic people counting during a demonstration for surveillance purposes~\citep{sanchez2020revisiting}.

Bringing an AI model to a real-world application is not immediate, as it implies the effort of integrating it in a functional system, including the necessary infrastructure, user interfaces, data pipelines, and other components required for the application to operate effectively in a production environment~\citep{hupont2022landscape}. Further, it is important to consider in the process the \textit{use cases} or variety of scenarios where the resulting system can be deployed. Use cases illustrate how users can utilize the AI system to accomplish their goals and therefore provide a key user-centric perspective on its functionality.

The European AI Act supports precisely this human-centric approach, putting the concept of \textit{intended purpose} at the centre of regulation~\citep{AIact,XAI_2023}. This paper discusses the AI Act as proposed by the Commission in April 2021~\citep{AIact}. We also mention some modifications made by the Council when adopting its common position (``general approach'') in December 2022~\citep{AIact_general}. The proposal is currently being debated by the EU co-legislators: the European Parliament and the Council and therefore the content of the final legislation may differ from what is described herein. The AI Act defines the \textit{intended purpose} of an AI system as: \\

{\it ``[...] the use for which an AI system is intended by the provider, including the specific context and conditions of use, as specified in the information supplied by the provider in the instructions for use, promotional or sales materials and statements, as well as in the technical documentation''} \\

\begin{figure}
\centering
  \includegraphics[width=0.7\textwidth]{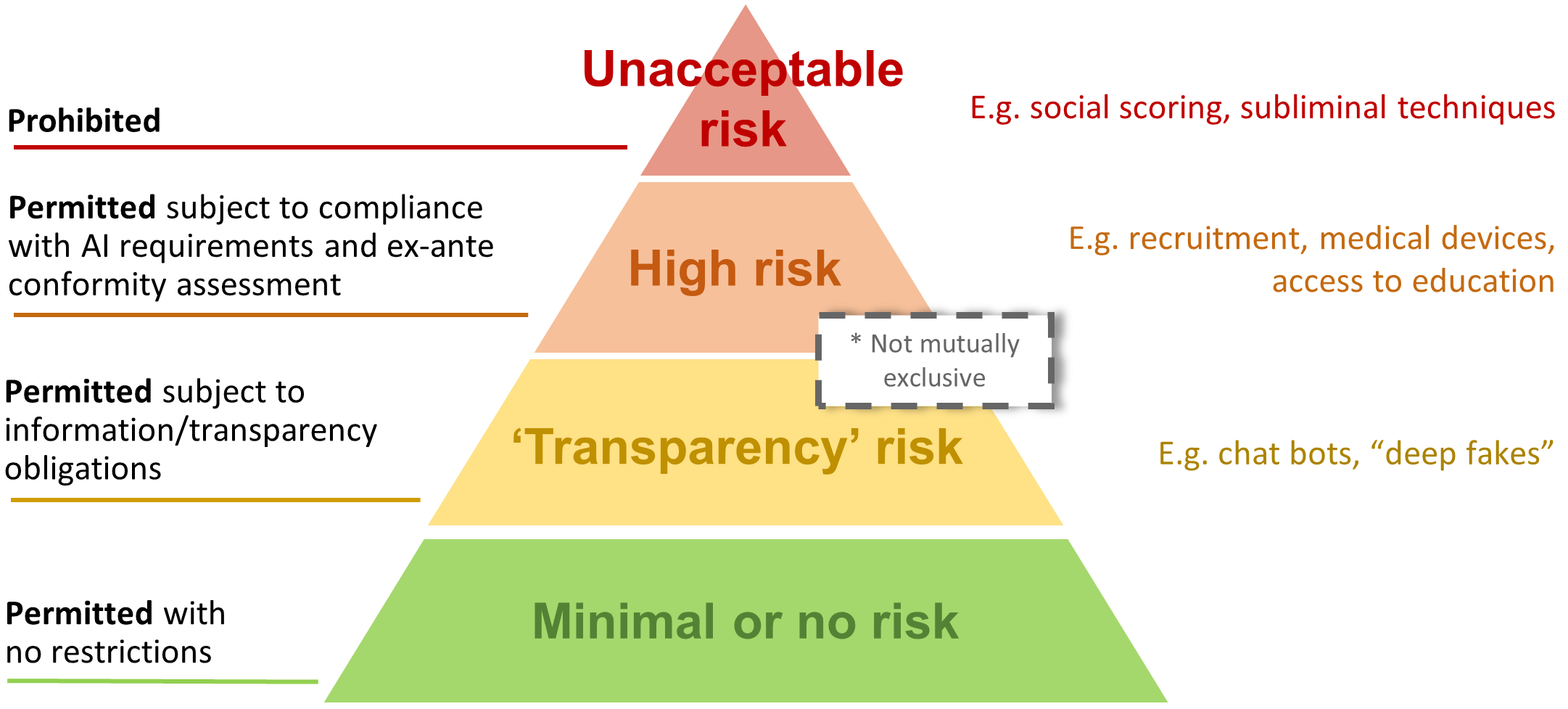}
  \caption{Risk level approach proposed in the AI Act.}
  \label{fig:risk_levels}
\end{figure}

According to the proposed regulation, the system's intended purpose determines its risk profile which can be, from highest to lowest: (1) \textit{unacceptable risk}, covering harmful uses of AI or uses that contradict European values; (2) \textit{high-risk}, covering uses identified through a list of high-risk application areas that may create an adverse impact on safety and fundamental rights; (3) \textit{transparency risk}, covering uses that pose risks of manipulation and are subject to a set of transparency rules (e.g. systems that interact with humans such as conversational agents, are used to detect emotions or generate or manipulate content such as \textit{deep fakes}); and (4) \textit{minimal risk}, covering all other AI systems. Figure~\ref{fig:risk_levels} illustrates this risk level approach. 

The AI Act establishes a clear set of harmonised rules that link use cases to risk levels, which in turn imply different legal requirements. AI systems classified as \textit{high-risk} according to these rules are those subject to conformity obligations. 
The rules to categorise an AI system's risk level depend on a series of key information elements that are essential to document its \textit{intended purpose}. We have compiled them in the list presented in Table~\ref{tab:info_elements_aia}. As can be seen, the system shall be put into context by providing  information on: the operational, geographical, behavioural and functional contexts of use that are foreseen; who will be the users and impacted stakeholders; and which are the system's inputs and outputs. In addition, it is as important to clearly specify the intended use of the system as its foreseeable potential misuses. Finally, there are three elements that are particularly important when it comes to identify an AI system's risk level. The first one is \textit{type of product}; AI Act's Annex II lists a number of Union product regulations (e.g. machinery, toys, medical devices regulations) and if the system --either as a component of the product or a product itself-- is subject to any of them, it is considered high-risk. The second element is \textit{safety component}; if an AI system is a safety component of a product or system then it is high-risk. The third one is \textit{application area}; AI Act's Annex III provides a concrete list of application areas under which an AI system is deemed high-risk (e.g. remote biometric identification systems, AI systems used to prioritise the dispatch of emergency services, those used as polygraphs by law enforcement).

Having all these information elements adequately covered in a unique use case documentation methodology would be a valuable tool both for policy makers and AI systems' providers to better navigate the AI Act and properly assessing AI systems' risk level as well as tailoring the different requirements. However, current AI documentation approaches fail to provide a full coverage as we will see in the next section.

\newcolumntype{E}{>{\raggedright\arraybackslash}m{0.15\textwidth}}
\newcolumntype{D}{>{\raggedright\arraybackslash}m{0.6\textwidth}}
\newcolumntype{R}{>{\raggedright\arraybackslash}m{0.15\textwidth}}

\begin{table}[h]
\centering
\footnotesize
\begin{tabular}{EDR}
\textbf{Element} & \textbf{Description}  & \textbf{Related legal text} \\
\toprule 
Intended purpose & 
Use for which an AI system is intended by the provider, including the specific context and conditions of use. 
& Art. 3(12)   \\
\midrule
User & Any natural or legal person, public authority, agency or other body, under whose authority the system is used. & Art. 3(4) \\
\midrule
Stakeholders  & Persons or group of persons on which the system is intended to be used and/or that are impacted by the AI system. & Aticle 7, Annex IV(2b) \\
\midrule
Input data  & Data provided to or directly acquired by the system on the basis of which the system produces an output. & Art. 3(32) \\
\midrule
Outputs  & Expected outputs of the AI system. & Art. 3(32), Art. 13(3vi) \\
\midrule
Foreseeable misuse  & Use of an AI system in a way that is not in accordance with its intended purpose, but which may result from reasonably foreseeable human behaviour or interaction with other systems.    
 & Art. 3(13) \\
\midrule
Type of product &  Type of product or service of which the AI system is a component or the product itself. It can be a machine (e.g. industrial machine, robot, motor vehicle), device (e.g. sensor, medical device), some other hardware (e.g. equipment) or a software (e.g. standalone application, software service). & Article 6, Annex II\\
\midrule
Safety component  &  Component of a product or of a system which fulfils a safety function for that product or system or the failure or malfunctioning of which endangers the health and safety of persons or property. & Art. 3(14) \\
\midrule
Application area & Area in which the AI system is intended to be applied (e.g. law enforcement, employment, marketing, education, healthcare). & Article 6, Annex III\\
\bottomrule
\end{tabular}
\caption{Key information elements related to use cases under the AI Act.} 
\label{tab:info_elements_aia}
\end{table}

\subsection{Existing approaches for AI documentation}

In the recent years key academic, government and industry players have proposed methodologies aiming at defining documentation approaches that increase transparency and trust in AI. Table~\ref{tab:sota} summarises the most popular ones, and analyses the extent to which they cover the use case-related information needs identified in the previous section. Note that the table exclusively considers documentation methodologies focusing on AI models, systems or services. For instance, it does not include works tackling only dataset documentation such as \textit{Datasheets for Datasets}~\citep{gebru2018datasheets}, \textit{The Dataset Nutrition Label}~\citep{chmielinski2022dataset} or \textit{Data Cards}~\citep{pushkarna2022data}.

Firstly, the table shows the importance the AI community is paying to documentation, as big tech (Google, IBM, Microsoft, Meta) and high stakes institutions such as the Organisation for Economic Co-operation and Development (OECD) are behind most adopted methodologies. For instance, Google's \textit{Model cards}~\citep{mitchell2019model} can now be automatically generated from the widely used TensorFlow framework\footnote{TensorFlow machine learning framework. Available at: \url{https://www.tensorflow.org/}}, which is strongly fostering its adoption by AI practitioners.

Nevertheless, as anticipated in the Introduction, the majority of methodologies have a strong technical focus. They have been generally conceived as tools for AI developers and providers to demonstrate AI models' performance and accuracy. Most recently proposed methodologies, including the \textit{Framework for the classification of AI systems} by the OECD~\citep{OECD}, \textit{AI usage cards}~\citep{wahle2023ai} and \textit{System cards}~\citep{metaSC}, are broadening towards other audiences such as policy-makers and end-users. Even though some methodologies do explicitly ask about the intended use of AI the system (e.g. \textit{``What is the intended use of the service output?"} in~\cite{arnold2019factsheets}, \textit{``Intended Use"} section in~\cite{mitchell2019model} and \textit{``Task(s) of the system''} in~\cite{OECD}), it is just in very broad terms and provided examples lack sufficient details to address complex legal concerns. Moreover, none of these methodologies are based on a formal standard or specification. 
In summary, to date there is no unified and comprehensive AI documentation approach focusing exclusively on use cases and covering information elements such as \textit{type of product}, \textit{safety component} and \textit{application area}. Our proposed \textit{use case cards} aim at bridging this gap.
 
\newcolumntype{F}{>{\raggedright\arraybackslash}m{.14\textwidth}}
\newcolumntype{N}{>{\centering\arraybackslash}m{.14\textwidth}}
\newcommand{\STAB}[1]{\begin{tabular}{@{}c@{}}#1\end{tabular}}
\begin{table}[h]
\centering\settowidth\rotheadsize{Largest text goes here                             }
\scalebox{0.59}{
\begin{tabular}{cFNNNNNNNN}
&  & \rotcell{\textbf{Model cards} \\ \cite{mitchell2019model}} & \rotcell{\textbf{AI FactSheets} \\ \cite{arnold2019factsheets}} & \rotcell{\textbf{AI fairness checklist} \\ \cite{madaio2020co}} & \rotcell{\textbf{Method cards} \\ \cite{adkins2022prescriptive}} & \rotcell{\textbf{OECD framework for AI systems} \\ \cite{OECD}}  & \rotcell{\textbf{AI usage cards} \\ \cite{wahle2023ai}} & \rotcell{\textbf{System cards} \\ \cite{metaSC}} & \rotcell{\textbf{Use case cards}} \\ 
\midrule
\multicolumn{2}{l}{\textbf{Proponent}}   &  Google & IBM & Microsoft & Meta & OECD &  Academia & Meta  & This work  \\ 
\midrule
\multicolumn{2}{l}{\textbf{Scope}}   & AI model   & AI service   & AI system  & AI model & AI system & AI system & AI system & Use case  \\ 
\midrule
\multicolumn{2}{l}{\textbf{Type of approach}}    & Information sheet     & Questionnaire    & Checklist     & Information sheet     & Questionnaire & Information sheet  & Interactive web page & Information sheet     \\ 
\midrule
\multicolumn{2}{l}{\textbf{Target stakeholders}}  & AI developers   & AI service providers  & AI developers   & AI developers    & Regulators, Society & AI researchers, developers &  End-users & Regulators, Society    \\ 
\midrule
\multicolumn{2}{l}{\textbf{Technical focus}}  & High    & High  & High  &  High   & High & High & Medium & Low \\ 
\midrule
\multicolumn{2}{l}{\textbf{Based on standard}}  & No   & No  & No  &  No   & No & No  & No & Yes \\ 
\midrule
\multirow{20}{*}{\STAB{\rotatebox[origin=c]{90}{\textbf{INFORMATION ELEMENTS}}}} & \textbf{Intended purpose}  & \CirclOK  & \CirclOK  & \CirclOK  & \CirclOK  & \RectTEC & \RectTEC & \CirclOK & \CirclOK  \\ 
\cmidrule{2-10}
& \textbf{Context of use}  & $\times$  & \RectTEC  & \CirclOK  & $\times$  & $\times$ & $\times$ & \CirclOK & \CirclOK  \\ 
\cmidrule{2-10}
& \textbf{User}                 & \CirclOK  & \CirclOK  & \CirclOK  & $\times$  & \CirclOK & $\times$ & \CirclOK & \CirclOK \\  
\cmidrule{2-10}
& \textbf{Stakeholders}         & $\times$  & $\times$  & \CirclOK  & $\times$  & \CirclOK & $\times$ & \CirclOK & \CirclOK \\  
\cmidrule{2-10}
& \textbf{Input data}               & \CirclOK  & \RectTEC  & $\times$  & \RectTEC  & \RectTEC & \CirclOK & \CirclOK & \CirclOK \\ 
\cmidrule{2-10}
& \textbf{Outputs}              & \CirclOK  & \RectTEC  & $\times$  & \RectTEC  & $\times$ & \CirclOK  & \CirclOK & \CirclOK \\ 
\cmidrule{2-10}
& \textbf{Foreseeable misuses}  & \CirclOK  & \CirclOK  & \CirclOK  & \RectTEC  & \CirclOK & \CirclOK  & $\times$ & \CirclOK \\ 
\cmidrule{2-10}
& \textbf{Type of product}      & $\times$  & $\times$  & $\times$  & $\times$  & $\times$ & $\times$ & $\times$ & \CirclOK \\ 
\cmidrule{2-10}
& \textbf{Safety component}     & $\times$  & $\times$  & $\times$  & $\times$  & $\times$ & $\times$ & $\times$ & \CirclOK \\  
\cmidrule{2-10}
& \textbf{Application area}    & $\times$  & \CirclOK  & $\times$  & $\times$  & \CirclOK & $\times$ & $\times$ & \CirclOK \\ 
\bottomrule
\end{tabular}
}
\caption{Comparison of state-of-the-art AI documentation approaches to our proposed \textit{use case cards}. The symbol \CirclOK \hspace{0.2mm}  denotes a good coverage of the information element, \RectTEC \hspace{0.2mm}  is used for elements only covered from a technical perspective, and $\times$ means no coverage. The methods have been assessed based on examples publicly available.}
\label{tab:sota}
\end{table}

\section{The \textit{use case card} documentation approach}
\label{sec:methodology}

\subsection{Revisiting UML for AI use case documentation}
\label{sec:revisit_uml}

\begin{figure}
\centering
  \includegraphics[width=\textwidth]{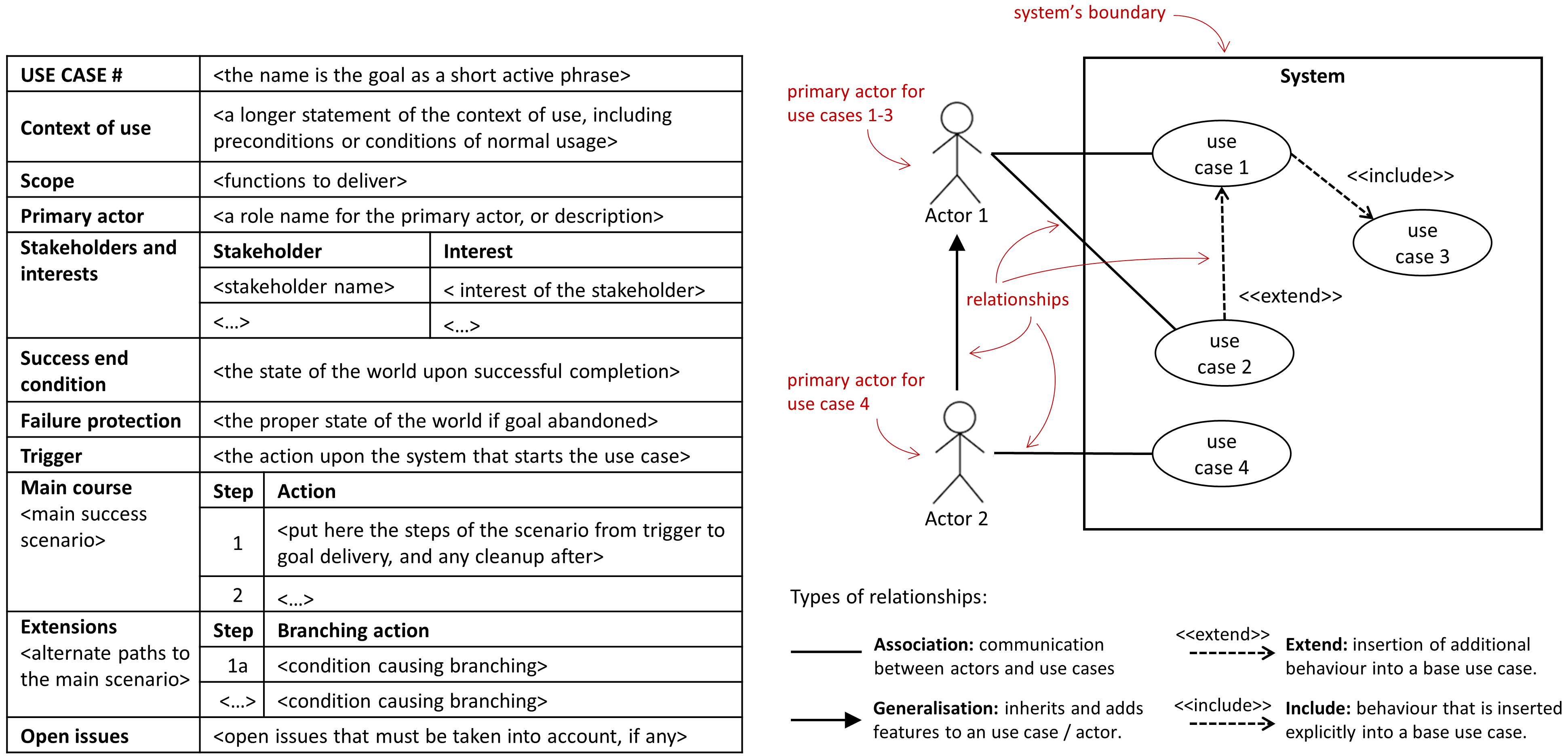}
  \caption{Traditional components of a use case modelled with UML. Left: table for use case description as proposed by~\cite{cockburn2001writing}. Right: visual elements, as established in the UML standard~\citep{UML251}. }
  \label{fig:UML_classic}
\end{figure}

Among use case modelling methodologies, the one proposed in the Unified Modelling Language (UML) specification is the most popular one in software engineering~\citep{kocc2021uml}. It has the advantage of being an official standard with more than 25 years of life and backed by a strong community~\citep{UML251}. Further, it is easy to use, offering a highly intuitive and visual way of modelling use cases by means of diagrams and a set of simple graphic elements (Figure~\ref{fig:UML_classic}). 

UML use cases capture what a system is supposed to do without entering into technical details (e.g. concrete implementation details, algorithm architectures). They rather focus on the context of use, the main actors using the system, and actor-actor and actor-system interactions. A use case is triggered by an \textit{actor} (it might be a person or group of persons), who is called \textit{primary actor}. The use case describes the various sets of interactions that can occur between the various actors, while the primary actor is in pursuit of a goal. A use case is completed successfully when the goal that is associated with it is reached. Use case descriptions also include possible extensions to this sequence, e.g., alternative sequences that may also satisfy the goal, as well as sequences that may lead to failure in completing the goal.

Once the use case has been modelled in a diagrammatic form (Figure~\ref{fig:UML_classic}-right), the next step is to describe it in a brief and structured written manner. The UML standard does not impose this step to be implemented, but it is commonly done in the form of a table. The most widely-used layout is the one proposed in~\cite{cockburn2001writing} and shown in Figure~\ref{fig:UML_classic}-left. 

The information elements related to use cases under the AI Act (c.f. Table~\ref{tab:info_elements_aia}) were found to closely match those of the software use case documentation under UML, e.g.: context of use and scope $\longleftrightarrow$ intended purpose; primary actor $\longleftrightarrow$ user; stakeholders and interests $\longleftrightarrow$ stakeholders; open issues $\longleftrightarrow$ foreseeable misuses; and main course $\longleftrightarrow$ inputs/outputs. For this reason, we decided to ground our proposed \textit{use case cards} in UML. The process of transforming classic UML use case diagrams into \textit{use case cards} was carried out in a co-design workshop with stakeholders that is detailed farther in Section~\ref{ssec:co-design}. In the next sections, we focus on presenting the final \textit{use case card} design and explaining how to fill it.

\subsection{Use case cards}
\label{sec:ucc}

\begin{figure}
\centering
  \includegraphics[width=\textwidth]{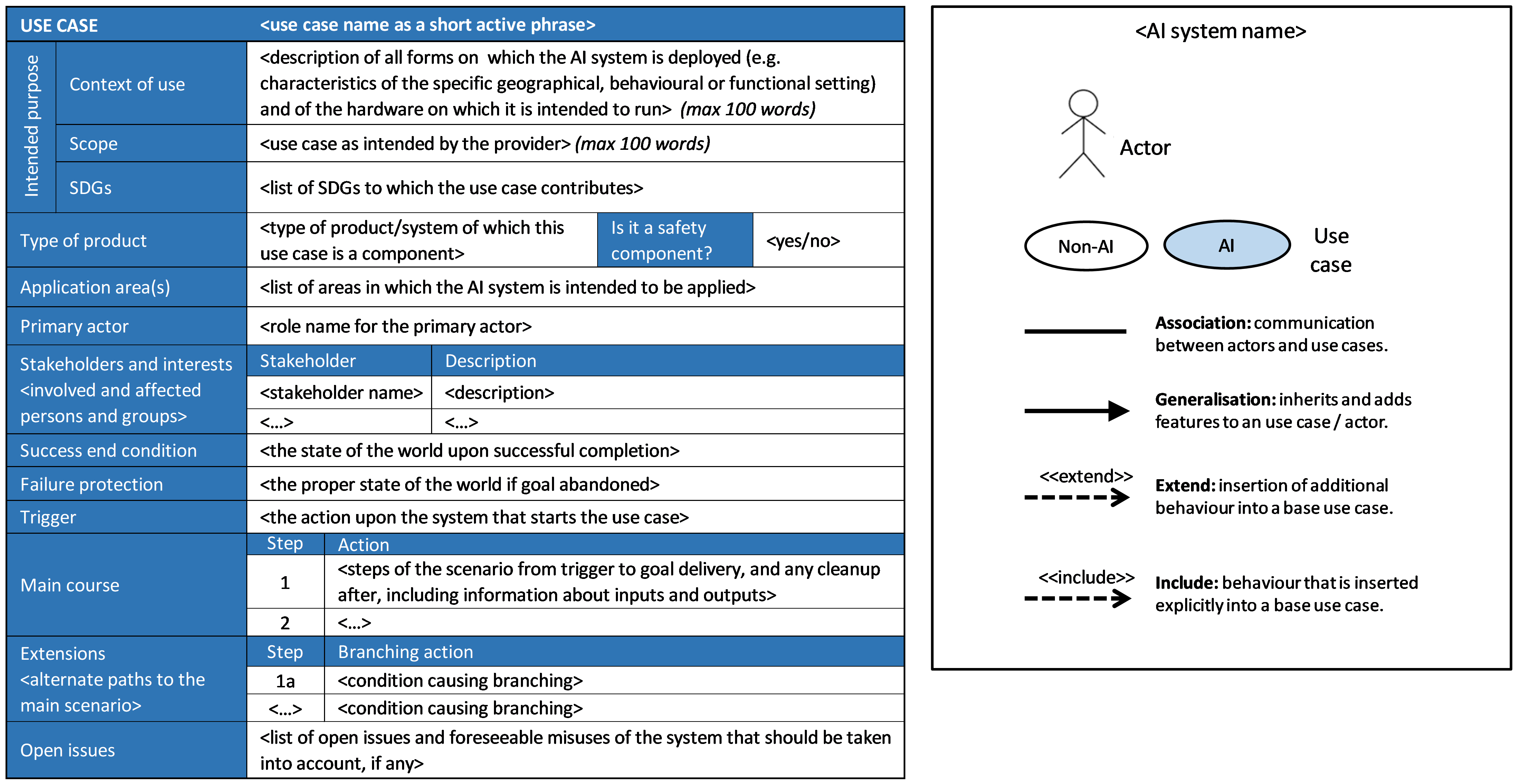}
  \caption{Proposed \textit{use case card} template. Left: use case table. Right: canvas for the visual modelling of the use case in the context of the AI system it belongs to or it is a component of.}
  \label{fig:template}
\end{figure}

The designed \textit{use case card} template is shown in Figure~\ref{fig:template}. It is composed of two main parts: a canvas for visual modelling (right) and an associated table for written descriptions (left). Both are very close to the UML standard, with only some few extra information elements inspired by European AI policies as follows. 
The canvas contains the following visual elements: 

\begin{itemize}
    \item \textbf{AI system boundary:} It delimits the functionalities of the AI system. It is represented by a rectangle that encloses all the use cases. 
    \item \textbf{Actors:} They represent users or external systems that interact with the AI system. They are depicted as stick figures placed outside the AI system's boundary. Actors can be individuals, groups, other software systems or even hardware devices. Each actor has a unique name to identify their role.
    \item \textbf{Use Cases:} They represent specific functionalities or behaviors of the AI system. They describe the interactions between actors and the AI system to achieve a specific goal. Use cases are represented as ovals within the system boundary. Differently from traditional UML, we distinguish between \textit{AI use cases} (with blue background) and \textit{non-AI use cases} (with white background). Each use case has a name that reflects the action or functionality it represents.
    \item \textbf{Relationships:} They show the associations and dependencies between actors and use cases. Associations are depicted by solid lines connecting an actor to a use case, indicating that the actor interacts with or participates in that particular use case. Associations can also exist between use cases to represent dependencies between different functionalities. ``Include'' and ``extend'' relationships are depicted with dashed arrows. ``Include'' shows that one use case includes the functionality of another use case. ``Extend'' indicates that a use case can extend another one with additional behavior. Generalization is depicted by a solid arrow pointing from the specialized actor to the generalized actor (i.e. the specialized actor inherits the characteristics and interactions of the generalized actor).
\end{itemize}

It is particularly important to understand the distinction between \textit{AI system} and \textit{use case}. The system perspective considers the AI system as a whole and helps in understanding its components (both AI and non-AI) and their relationships. \textit{Use cases}, on the other hand, represent the specific interactions that actors have with the system and the functionalities the system provides them. By distinguishing systems from use cases, UML provides a modular and flexible modelling approach, allowing to focus on different aspects of the system at different levels of abstraction and granularity. 
Also note that for a system to be considered \textit{AI system} in a \textit{use case card} it has to content at least one \textit{AI use case}. 

The table layout has some changes with respect to the one proposed in~\citep{cockburn2001writing}. First, the \textit{intended purpose} of the system encompasses three fields. Two of them already appeared in the original table, namely, \textit{context of use} and \textit{scope}. Both are to be filled with a short text description; we recommend a maximum of 100 words. Remaining field is \textit{Sustainable Development Goals (SDGs)} and its values should by picked from the official United Nations' list presented in Appendix A's Figure~\ref{fig:SDGs}. Note that the purpose of this field is stating the SDGs to which the use case contributes (i.e. has a positive impact). 

In addition, three new fields have been added as they are essential to determine the use case's risk level --and thus the one of the AI system containing it-- according to the AI Act. Their description can be found in Table~\ref{tab:info_elements_aia} and below we comment on their possible values:

\begin{itemize}
    \item \textbf{Type of product:} It must be one value from the list in Appendix A's Table~\ref{tab:products}. Top rows in the list correspond to type of products that might be subject to other EU regulations and, as such, be high-risk according to AI Act's Annex II.
    \item \textbf{Is it a safety component?:} This ``yes/no'' field determines whether the use case fulfills a safety function for a product or system whose failure might harm persons or material. It is therefore a flag field that indicates a high-risk level.
    \item \textbf{Application area(s):} One or more areas of application of the use case, as listed in Appendix A's Table~\ref{tab:areas}. Some of these areas are high-risk under the AI Act and therefore need to be clearly identified.
\end{itemize}

Remaining fields correspond one-to-one with to those in the original table. The only change appears in the description of the \textit{open issues} field where we have emphasised the need to include   \textit{foreseeable misuses} of the system.

\subsection{Filling in use case cards}
\label{sec:filling_ucc}

This section illustrates the process of filling in a \textit{use case card} through the example of a scene narrator application installed in a smartphone. This AI-based application aims at helping people with visual impairments to obtain information about their environment, namely, about surrounding objects, text (e.g. panels, signs, menus) and people (both familiar and unknown persons). The user wears goggles connected to the smartphone, allowing to take a picture of the scene by pressing a button in the right ear temple. Then, the application narrates with a synthetic voice and in natural language the scene description, such as:\\

\textit{``You are in an office; there are four persons in front of you, the one on your left is John; there is a table with four chairs and the exit door is at the end of the room on the left hand side.''} \\

This application is inspired by real products in the market, including Microsoft's \textit{Seeing AI App}~\citep{narrator_1}, Cloudsight's \textit{TapTapSee}~\citep{narrator_2} and Google's \textit{Lookout}~\citep{narrator_3}. It is a complex application in computational terms, as it combines AI algorithms of different nature: object and person detection, optical character recognition (OCR), face recognition, text and synthetic voice generation. There are also data use and data privacy issues to be carefully addressed, e.g., regarding the management of captured facial images or the possibility of using extracted scene information for other purposes that assisting visually impaired such as targeted marketing.

\begin{figure}
\centering
  \includegraphics[width=\textwidth]{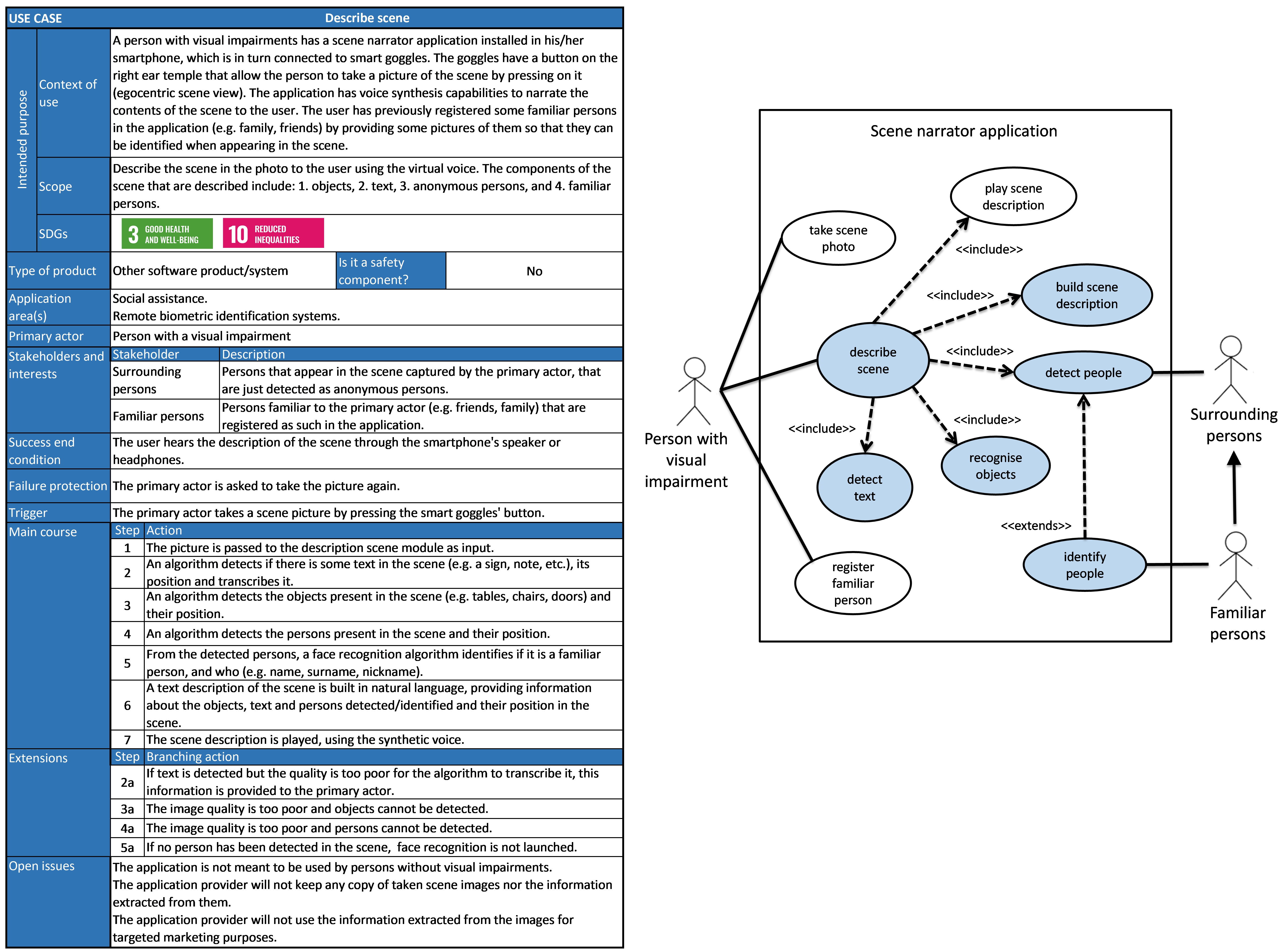}
  \caption{Filling in a \textit{use case card}: example of a scene narrator application.}
  \label{fig:scene_narrator_ucc}
\end{figure}

We propose the \textit{use case card} presented in Figure~\ref{fig:scene_narrator_ucc}. First, we focus on the visual modelling side. The key questions to pose is what is the AI system, which are the use cases within it we want to document and the main actors involved. The AI system can be easily identified as the \textit{scene narrator application}. This system may have multiple use cases, ranging from classic software functionalities (e.g. installing the app, user registration, user logging, manage settings) to the more complex AI-based functionalities related to the scene narration part (i.e. object/person detection, OCR, face recognition, etc.). We decide to include within the system's boundary only the uses cases directly linked to the scene narration functionality for the sake of clarity. 
Then, we reflect on a simplified interaction pipeline for the person with visual impairment to get a scene description, which is: \textit{opening the app in the smartphone $\rightarrow$ taking a picture of the scene $\rightarrow$ the system computes scene description $\rightarrow$ the person listens to the audio narration}.

Within this pipeline, we realise that the whole AI core is contained under the computation of scene description phase. We therefore decide to introduce a \textit{describe scene} use case as the principal one, which includes all AI-based functionalities (those with blue background colour). By modelling \textit{describe scene} as the main use case with ``include'' dependencies to other AI functionalities, we simplify the documentation process to a single UML table\footnote{Note that several use case tables can be linked to the same UML diagram, depending on how the system is modelled, how many components it has, and the level of granularity we want for documentation.}. We additionally decide to show some non-AI use cases in the diagram to provide a complete and self-contained overview of the pipeline, namely:   \textit{take scene photo} and \textit{register familiar person}. The \textit{register familiar person} use case is particularly interesting, as it shows that certain persons (e.g. family, friends, caregivers) might be registered in the platform by the user, and thus subject to identification through face recognition. 
The last point to define in the diagram are the actors involved. The main actor is clearly the person with visual impairments as s/he is the one triggering the scene narration process. The modelling process has nevertheless allowed to identify other relevant actors, namely the (unknown) surrounding persons that might appear in the scene and the familiar faces that might eventually be present. Note that  \textit{surrounding persons} are a generalisation of \textit{familiar persons}, and that the \textit{identify people} use case  ``extends'' the \textit{detect people} one.

After the visual modelling exercise, we proceed to complete the table associated to the main use case \textit{describe scene}. The \textit{context of use} field provides an overview of pre-conditions and conditions of normal usage (e.g. the app is already installed in the smartphone, the primary actor wears goggles, s/he has already registered some familiar faces in the system), while \textit{scope} delimits the concrete functionality of the use case. This use case has a strong positive social impact, allowing for a better inclusion and social life for the visually impaired, and therefore contributes to two SDGs: \textit{good health and well-being} and \textit{reduce inequalities}. The use case is part of a software product and may not be considered a safety component, as it is meant to assist but not to fulfil a safety function. Interestingly, it has two application areas. The first one is \textit{social assistance}, and the second one is \textit{remote biometric identification systems} as it includes face recognition to identify familiar people. This is particularly important as the former is not considered a high-risk application area under the AI Act, while the latter does. Therefore, if the system's provider prefers to bring the application to the market as a low-risk one, the face recognition functionality should be removed. 
The following fields are relatively straightforward to document, as they merely describe the main actors and course of actions within the use case. In our example, the \textit{main course} field contains as steps the calls to the different AI algorithms. \textit{Extensions} tackle problems that may arise, e.g. if the taken picture has poor quality, which are simply addressed with the \textit{failure protection} mechanism of asking the person to retake the shot. Last but of extremely importance, the \textit{open issues} field allows the provider to clearly state that the application is conceived for ethical use. It stresses that the system is not intended for use by people who are not visually impaired, clarifies that data privacy is adequately treated (the provider does not keep a copy of taken scene images) and that under no circumstances will the provider do any marketing with the extracted information.

Through this example, we have shown that {use case cards} is a powerful, standardised methodology to document AI use cases. Besides the end of documentation, the process of filling in a \textit{use case card} fosters reflections of the utmost importance about an AI system, such as its risk level, foreseeable misuses and failure protection mechanisms to put in place. 
Appendix~\ref{annex_B} provides four other \textit{use case cards} involving different types of AI systems with varying levels of complexity, to provide the reader with a variety of illustrative examples.

\section{Co-designing and validating \textit{use case cards} with key stakeholders}
\label{sec:user_study}

\begin{figure}
\centering
   \includegraphics[width=0.75\textwidth]{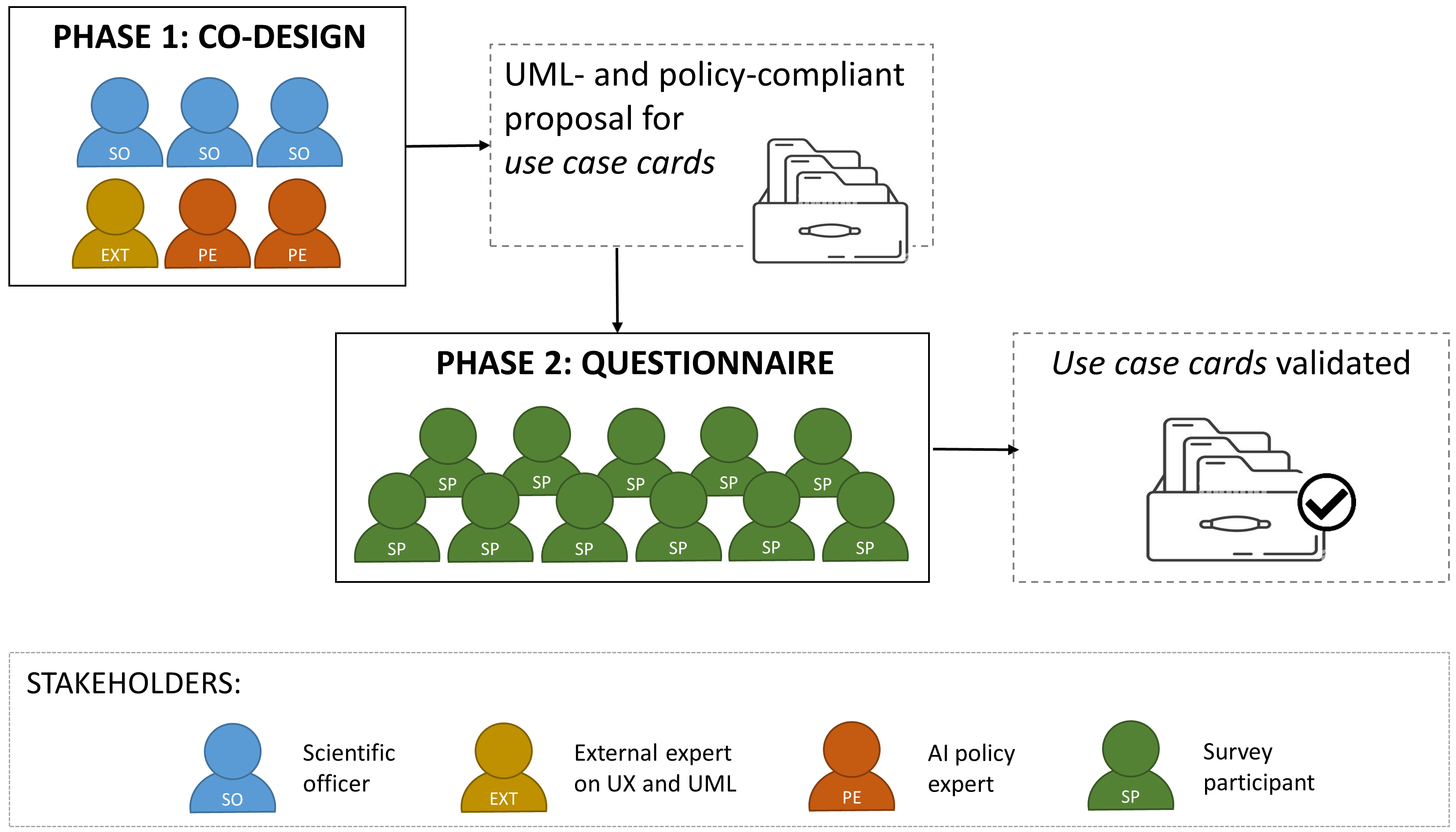}
   \caption{Two-phase protocol followed for the design and validation of \textit{use case cards} with key stakeholders.}
   \label{fig:protocol}
 \end{figure}

The \textit{use case card} methodology was developed following a two-phase protocol with key stakeholders, as depicted in Figure~\ref{fig:protocol}.
First, we carried out a co-design workshop involving two European Commission (EC) policy experts, three EC scientific officers and an external expert on User eXperience (UX) and UML. The resulting version of \textit{use case cards} was then evaluated in a second phase through a questionnaire to 11 scientists contributing to different EU digital policy initiatives, and with varying expertise levels on UML and the AI Act. In the following, we provide details on the implementation of both phases and present the main results.

\subsection{Co-design process}
\label{ssec:co-design}

Co-design, co-creation or participatory design refers to an approach where stakeholders come together, as equals, to conceptually develop  solutions that respond to certain matters of concern~\citep{zamenopoulos2018co}. As such, the co-design method aims to develop a solution ``with'' the target individuals/groups rather than ``for'' them. There has been a increasing trend in recent years towards greater inclusion of stakeholders in designing and carrying out research through adoption of co-design methods~\citep{nesbitt2022user}. Given the multidisciplinary nature of our work, involving both policy and technical matters, we decided to take advantage from this methodology in this first design phase.

The co-design phase involved six participants. Two of them are EC policy experts with legal background, and having a high involvement and proficiency expertise on the AI Act. Three are EC scientific officers with a proficiency background on AI and medium-to-high knowledge on UML. It is important to note that, although these three experts have primarily a technical profile, they are involved on a daily basis in digital policy issues, including scientific advice related to the AI Act. Finally, we invited an external expert with high expertise on AI, and a proficiency background on UX and UML.


We organised a two-day physical workshop to conduct the co-design of \textit{use case cards}. Scientific officers alternated in making questions and taking copious notes throughout the workshop, counting with all participants' permission.

The three scientific officers and the external UML/UX expert prepared a three-hour tutorial on UML to kick off the first day. The tutorial started with the presentation of the UML standard~\citep{UML251}, with particular emphasis on the use case modelling part.
Then, three exemplar AI use cases modelled in classic UML format (c.f. Figure~\ref{fig:UML_classic}) were presented for illustrative purposes: an affective music recommender, a driver monitoring system and a smart-shooting camera system. 

After the tutorial, the six participants engaged in a guided discussion covering the following key points: 

\begin{itemize}
    \item Potential of UML as a standard methodology for AI use case documentation.
    \item Relevance, clarity and adequateness of the UML diagram and related table with regard to the AI Act (e.g. missing fields, ease to understand/implement).
    \item Relevance of the method for the assessment of an AI system's risk level according to the AI Act.
\end{itemize}

Results can be summarised as follows. First, participants unanimously agreed on the high overlap between UML's information elements and those required to document use cases under the AI Act (c.f. Table~\ref{tab:info_elements_aia}). Therefore the standard was considered fit for purpose. 
 Participants however identified missing fields essential in the context of the AI Act 
 and that should be added to the UML table, namely: (i) the type of product to which the AI system belongs; (ii) its application area(s); and (iii) whether the use case is a safety component of a product. 

Participants raised important additional points. They mentioned different uses of the methodology, including the creation of a public repository of AI use cases, useful in the context of the registration processed mentioned in Article 51 and 
Annex VIII--part II of the AI Act. 
This repository would be a valuable and usable tool to help companies --and more particularly SMEs, with more limited legal resources-- identify the risk level of their AI systems: \textit{``use case cards would give companies a hook to go through the AI Act''}. Authorities would also benefit from such repository, allowing to \textit{``have a better overview of the landscape of existing AI systems''} and \textit{``engage with companies to articulate bordercases''}. Although not an explicit information requirement under the AI Act, given its human-centered nature it was deemed interesting to include the link of each use case to Sustainable Development Goals (SDGs) which \textit{``would help keep track of AI-for-good applications''}.

During the second workshop day, participants proceeded to the design of \textit{use case cards} according to the findings identified the previous day. They first added the four missing fields (i.e. ``type of product'', ``application area(s)'', ``is it a safety component?'' and ``SDGs'') to the UML table, and agreed on its final layout (e.g. colors, order/position of the different fields). Then they developed altogether the list of type of products (c.f. Appendix A's Table~\ref{tab:products}) and application areas (c.f. Appendix A's Table~\ref{tab:areas}), carefully considering AI Act's Annex II and III, respectively. Finally, participants concluded with a practical exercise, where they converted the three UML use cases in the tutorial to the new \textit{use case card} format. They additionally implemented two new \textit{use case cards}: the scene narrator one (presented in the previous section) and a student proctoring one. 
New \textit{use case cards} can be found in Appendix~\ref{annex_B}. This final exercise allowed us to confirm the ease of use and implementation of the methodology, whose adaptation is \textit{``straightforward with respect to traditional UML''} as confirmed by the UML/UX expert.

\subsection{Questionnaire-based validation study}

Once the first solid version of the \textit{use case cards} was available, we conducted a questionnaire-based study to validate two main aspects. On the one hand, those components referring to the clarity and complexity of the proposed approach, such as its learning curve, its level of detail and granularity, the importance of the visual components with respect to the table, as well as open questions regarding possible missing or unnecessary fields. On the other hand, those elements related to the level of contextualization with respect to the AI Act, risk level assessment, requirements, etc. A summary of the questions is provided in Table \ref{tab:questions}. As can be seen, 9 questions were designed to have a possible answer aligned with a 5-point Likert scale, 2 questions allowed for a yes/no answer plus an elaboration if the answer was yes, and 2 questions were designed as completely open questions. 

\newcolumntype{Q}{>{\raggedright\arraybackslash}m{0.04\textwidth}}
\newcolumntype{U}{>{\raggedright\arraybackslash}m{0.78\textwidth}}
\newcolumntype{a}{>{\raggedright\arraybackslash}p{0.15\textwidth}}
\newcolumntype{b}{>{\raggedright\arraybackslash}p{0.18\textwidth}}
\begin{table}
\caption{Summary of the questionnaire. Qx denotes 5-point Likert-scale questions and OQx stands for open questions.}
\footnotesize
\centering
\scalebox{0.91}{
\begin{tabular}{Qaaaab}
\toprule
\multicolumn{6}{l}{\textbf{Questions and possible possible answers}}\\
\toprule
        Q1 & \multicolumn{5}{U}{Level of expertise on the AI Act:} \\
        & \CircleLikert \hspace{0.2mm} {\tiny None} & \CircleLikert \hspace{0.2mm} {\tiny Low} & \CircleLikert \hspace{0.2mm} {\tiny Mid} & \CircleLikert \hspace{0.2mm} {\tiny High} & \CircleLikert \hspace{0.2mm} {\tiny Very high}  \\
        \cmidrule{2-6}    
        Q2 & \multicolumn{5}{U}{Level of expertise on UML:} \\
        & \CircleLikert \hspace{0.2mm}  {\tiny None} & \CircleLikert \hspace{0.2mm}  {\tiny Low} & \CircleLikert \hspace{0.2mm}  {\tiny Mid} & \CircleLikert \hspace{0.2mm}  {\tiny High} & \CircleLikert \hspace{0.2mm}  {\tiny Very high} \\ 
        \toprule
        Q3 & \multicolumn{5}{U}{Difficulty to understand the use cases:} \\ 
        & \CircleLikert \hspace{0.2mm} {\tiny Very difficult} & \CircleLikert \hspace{0.2mm} {\tiny Somewhat \newline \CircleLikertFake \hspace{0.2mm}  difficult} & \CircleLikert \hspace{0.2mm} {\tiny Neutral} & \CircleLikert \hspace{0.2mm} {\tiny Somewhat \newline \CircleLikertFake \hspace{0.2mm} easy} & \CircleLikert \hspace{0.2mm} {\tiny Very easy} \\
        \cmidrule{2-6}         
        Q4 & \multicolumn{5}{U}{How would you rate the level of detail provided in the table?} \\
        & \CircleLikert \hspace{0.2mm} {\tiny Too little \newline \CircleLikertFake \hspace{0.2mm} detailed} & \CircleLikert \hspace{0.2mm} {\tiny Little detailed} & \CircleLikert \hspace{0.2mm} {\tiny Adequate} & \CircleLikert \hspace{0.2mm} {\tiny Quite detailed} & \CircleLikert \hspace{0.2mm} {\tiny Too detailed} \\
        \cmidrule{2-6}           
        Q5 & \multicolumn{5}{U}{How important do you consider the UML diagram with regard to the table for the use case?} \\
        & \CircleLikert \hspace{0.2mm} {\tiny Not important} & \CircleLikert \hspace{0.2mm} {\tiny Slightly \newline \CircleLikertFake \hspace{0.2mm} important} & \CircleLikert \hspace{0.2mm} {\tiny Moderately \newline \CircleLikertFake \hspace{0.2mm}  important} & \CircleLikert \hspace{0.2mm} {\tiny Important} & \CircleLikert \hspace{0.2mm} {\tiny Very \newline \CircleLikertFake \hspace{0.2mm} important} \\
        \cmidrule{2-6}           
        Q6 & \multicolumn{5}{U}{How do you assess the learning curve of the use case cards?} \\
        & \CircleLikert \hspace{0.2mm} {\tiny Not \newline \CircleLikertFake \hspace{0.2mm} appropriate} & \CircleLikert \hspace{0.2mm} {\tiny Slightly \newline \CircleLikertFake \hspace{0.2mm} appropriate} & \CircleLikert \hspace{0.2mm} {\tiny Moderately \newline \CircleLikertFake \hspace{0.2mm} appropriate} & \CircleLikert \hspace{0.2mm} {\tiny Quite \newline \CircleLikertFake \hspace{0.2mm} appropriate} & \CircleLikert \hspace{0.2mm} {\tiny Very \newline \CircleLikertFake \hspace{0.2mm}  appropriate} \\
        \cmidrule{2-6}    
        Q7 & \multicolumn{5}{U}{Is the \textit{use case card} well contextualised in relation to the AI Act?} \\
        & \CircleLikert \hspace{0.2mm} {\tiny Not at all} & \CircleLikert \hspace{0.2mm} {\tiny Very little} & \CircleLikert \hspace{0.2mm} {\tiny Neutral} & \CircleLikert \hspace{0.2mm} {\tiny Somewhat} & \CircleLikert \hspace{0.2mm} {\tiny To a great extent}\\
        \cmidrule{2-6}            
        Q8 & \multicolumn{5}{U}{Does the \textit{use case card} provide information to assess the risk-level according to the AI Act?} \\
        & \CircleLikert \hspace{0.2mm} {\tiny Not at all} & \CircleLikert \hspace{0.2mm} {\tiny Very little} & \CircleLikert \hspace{0.2mm} {\tiny Neutral} & \CircleLikert \hspace{0.2mm} {\tiny Somewhat} & \CircleLikert \hspace{0.2mm} {\tiny To a great extent}\\
        \cmidrule{2-6}            
        Q9 & \multicolumn{5}{U}{In the context of the AI Act, \textit{use case card} is appropriate for:  \newline  (1) risk-level assessment, (2) requirements, (3) catalogue of usages, (4) other:}\\
        & \CircleLikert \hspace{0.2mm} {\tiny Strongly \newline \CircleLikertFake \hspace{0.2mm} disagree} & \CircleLikert \hspace{0.2mm} {\tiny Somewhat \newline \CircleLikertFake \hspace{0.2mm} disagree} & \CircleLikert \hspace{0.2mm} {\tiny Not sure} & \CircleLikert \hspace{0.2mm} {\tiny Somewhat \newline \CircleLikertFake \hspace{0.2mm} agree} & \CircleLikert \hspace{0.2mm} {\tiny Strongly agree} \\
        \toprule
        OQ1 & \multicolumn{5}{U}{Is there any important field that you miss in the table?} \\
        & & \multicolumn{4}{U}{\CircleLikert \hspace{0.2mm} {\tiny Yes} \CircleLikert \hspace{0.2mm} {\tiny No; if Yes, please indicate which one}}\\
        \cmidrule{2-6}    
        OQ2 & \multicolumn{5}{U}{Is there any field that you would remove?} \\
        & & \multicolumn{4}{U}{\CircleLikert \hspace{0.2mm} {\tiny Yes} \CircleLikert \hspace{0.2mm} {\tiny No; if Yes, please indicate which one}}\\
        \cmidrule{2-6}    
        OQ3 & \multicolumn{5}{U}{Please specify other potential uses:}\\
        \cmidrule{2-6}    
        OQ4 & \multicolumn{5}{U}{Please insert here any additional comment you may have:}\\ 
\bottomrule
\end{tabular}
}
\label{tab:questions}   
\end{table}

The online survey included an introduction with the description of the project, the main goals and  procedure. Then a brief introduction of the main components of use cases modelled with UML was provided, followed by a short description of the proposed structure for the \textit{use case cards}. After some demographic questions, the participants were provided with three exemplar \textit{use case cards}. 
The first one corresponds to the scene narrator system previously presented in Section~\ref{sec:filling_ucc} (Figure~\ref{fig:scene_narrator_ucc}). Remaining two correspond to the driver attention monitoring system and the student proctoring tool presented in Appendix~\ref{annex_B} (Figures~\ref{fig:driver_monitoring_ucc} and~\ref{fig:proctoring_ucc}, respectively).

\begin{figure}
\centering
  \includegraphics[width=0.7\textwidth]{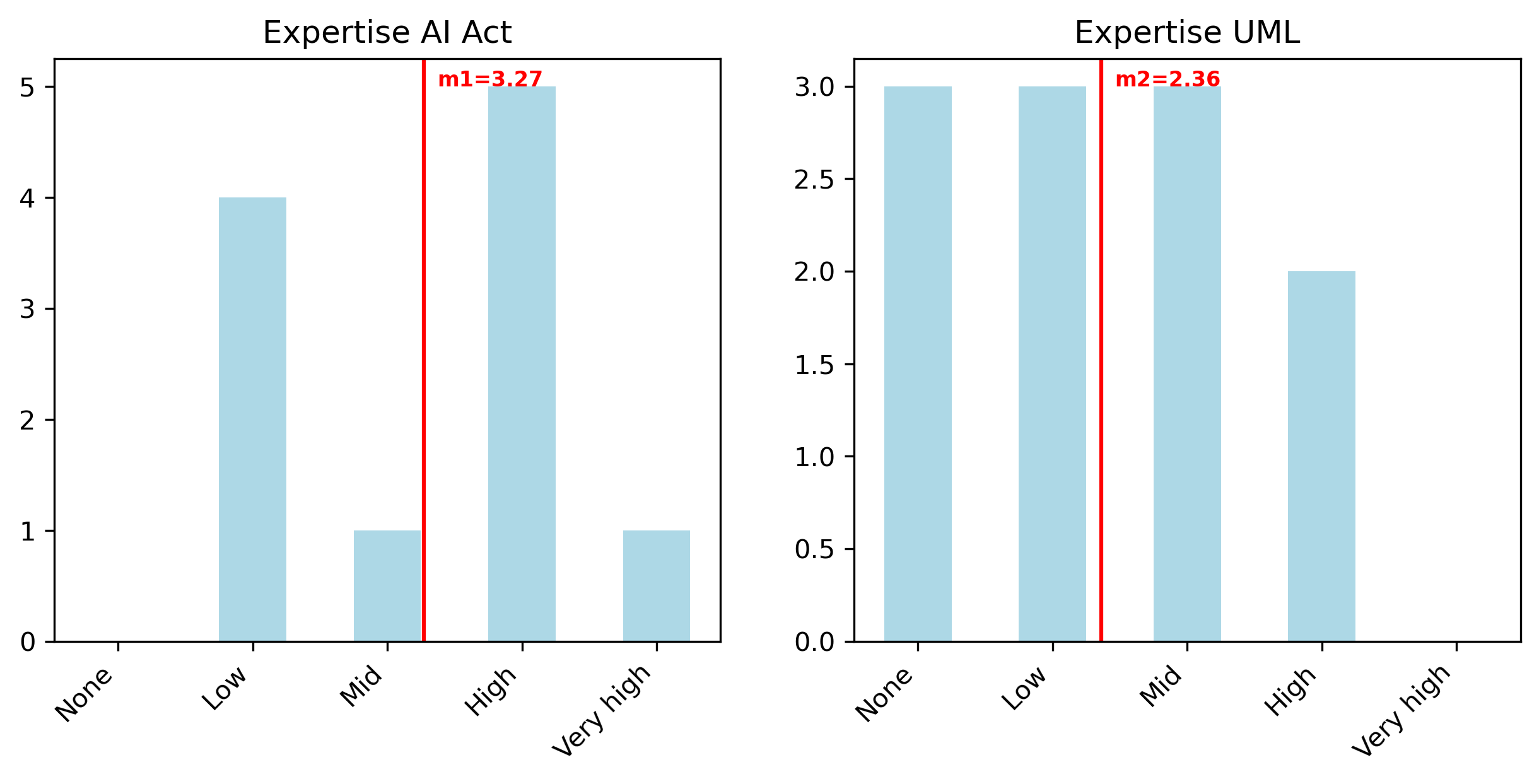}
  \caption{Histograms of the answers to questions 1 and 2, and mean values.}
  \label{fig:q1q2}
\end{figure}

We involved 11 participants (5 female, 5 male, 1 prefer not to say), 7 of whom had a technical background (computer scientists/engineers), and the rest with varied profiles including 1 legal expert, 1 social scientist and 1 mathematician. All of them had experience in trustworthy AI, science for policy, and the AI Act, as well as varying degrees of knowledge of UML. More specifically, their knowledge about the AI Act was self-assessed between “low” and “very high”, with mean $M1=3.27$ (question 1, Figure \ref{fig:q1q2}-left), whereas about UML it was self-assessed between “none” and high, with mean $M2=2.36$ (question 2,  Figure \ref{fig:q1q2}-right). Since the \textit{use case cards} are intended to be used in the context of the AI Act, it is coherent to validate them with participants with some knowledge of the AI Act. However, in principle, it is not strictly necessary to have knowledge of UML, so validation should incorporate participants with little or no knowledge of UML.

\begin{figure}
\centering
  \includegraphics[width=\textwidth]{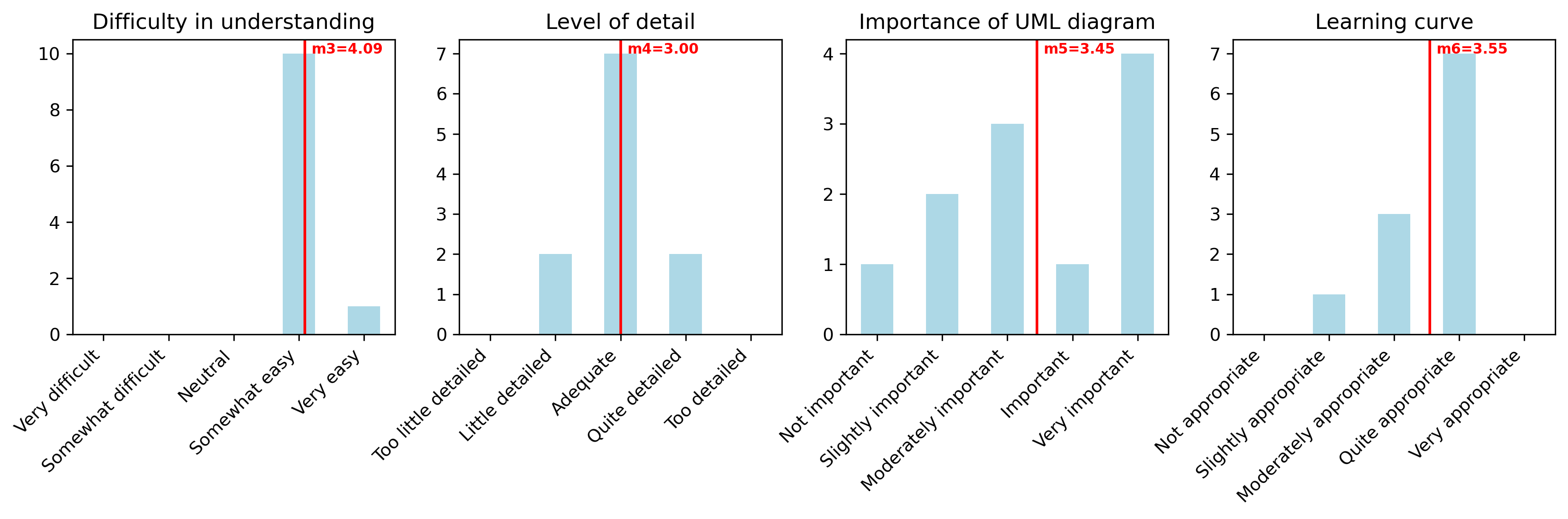}
  \caption{Histograms of the answers to questions 3 to 6, and mean values.}
  \label{fig:q3q6}
\end{figure}

\begin{figure}
\centering
  \includegraphics[width=0.7\textwidth]{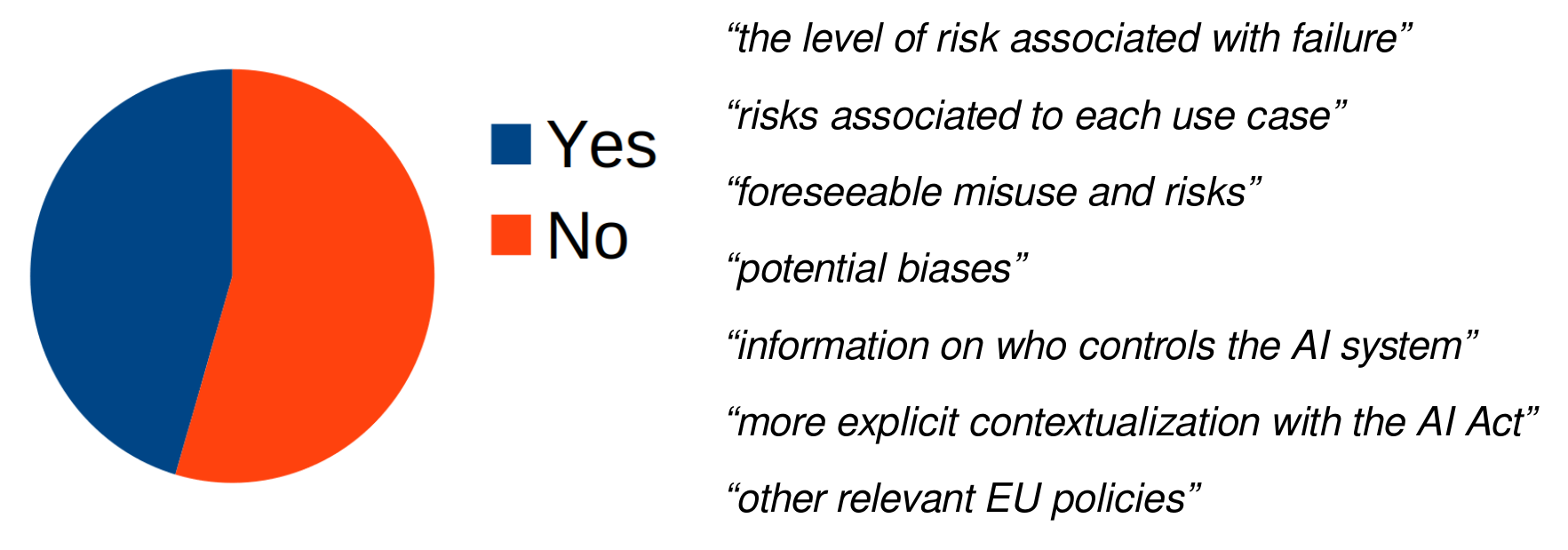}
  \caption{Answers to open question 1: \textit{``Is there any important field that you miss in the table?''}.}
  \label{fig:oq1}
\end{figure}

\begin{figure}
\centering
  \includegraphics[width=0.64\textwidth]{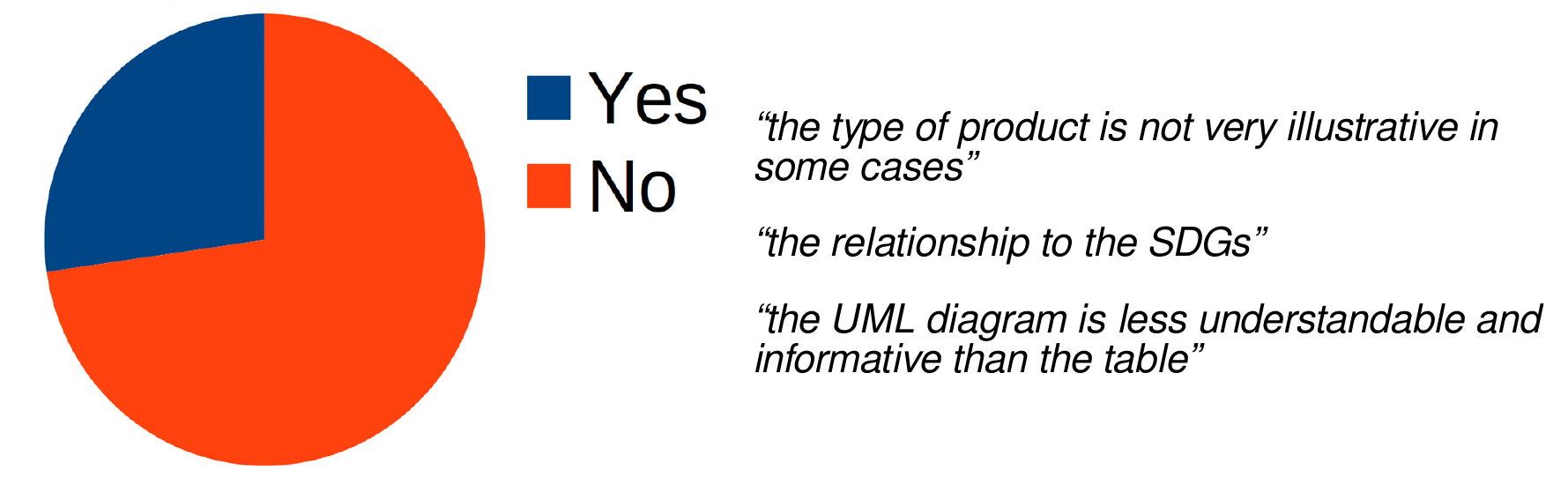}
  \caption{Answers to open question 2: \textit{``Is there any field that you would remove?''}.}
  \label{fig:oq2}
\end{figure}

Figure \ref{fig:q3q6} shows the histograms of answers for the questions related to the intrinsic features of the method. The difficulty to understand the three exemplar \textit{use case cards} was assessed as “somewhat easy” ($M3=4.09$), the level of detail as “adequate” ($M4=3.00$), the importance of the UML diagram (the canvas) between “moderately important” and “important” ($M5=3.45$), and the learning curve at the midpoint between “moderately appropriate” and “quite appropriate” ($M6=3.55$). Regarding the question on missing fields (OQ1), 6 participants answered “no” and 5 “yes”. The suggestions provided by those who answered “yes” can be seen in Figure~\ref{fig:oq1}. Most of them can be easily integrated into the “Open issues” field of the table. Other suggestions such as “\textit{more explicit contextualization with the AI Act}” or “\textit{other relevant EU policies}” could be considered in future versions. And as for the question on possible dispensable fields (OQ2), $73\%$ of the participants answered  “no”, and $27\%$ “yes”. As depicted in Figure \ref{fig:oq2}, there were three concerns, one referring to the type of product, another focusing on the Sustainable Development Goals (SDGs), and one comment on the UML diagram. First, it is important to note that the type of product has to be considered together with the specific area. Otherwise, we cannot obtain a detailed classification.  On the other hand, we believe that asking about the SDGs can have positive effects on AI systems providers, as a way for them to consider whether or not their systems contribute to sustainable development. Finally, the importance of the UML diagram has been positively assessed by most of the participants in question 5.

\begin{figure}
\centering
  \includegraphics[width=0.7\textwidth]{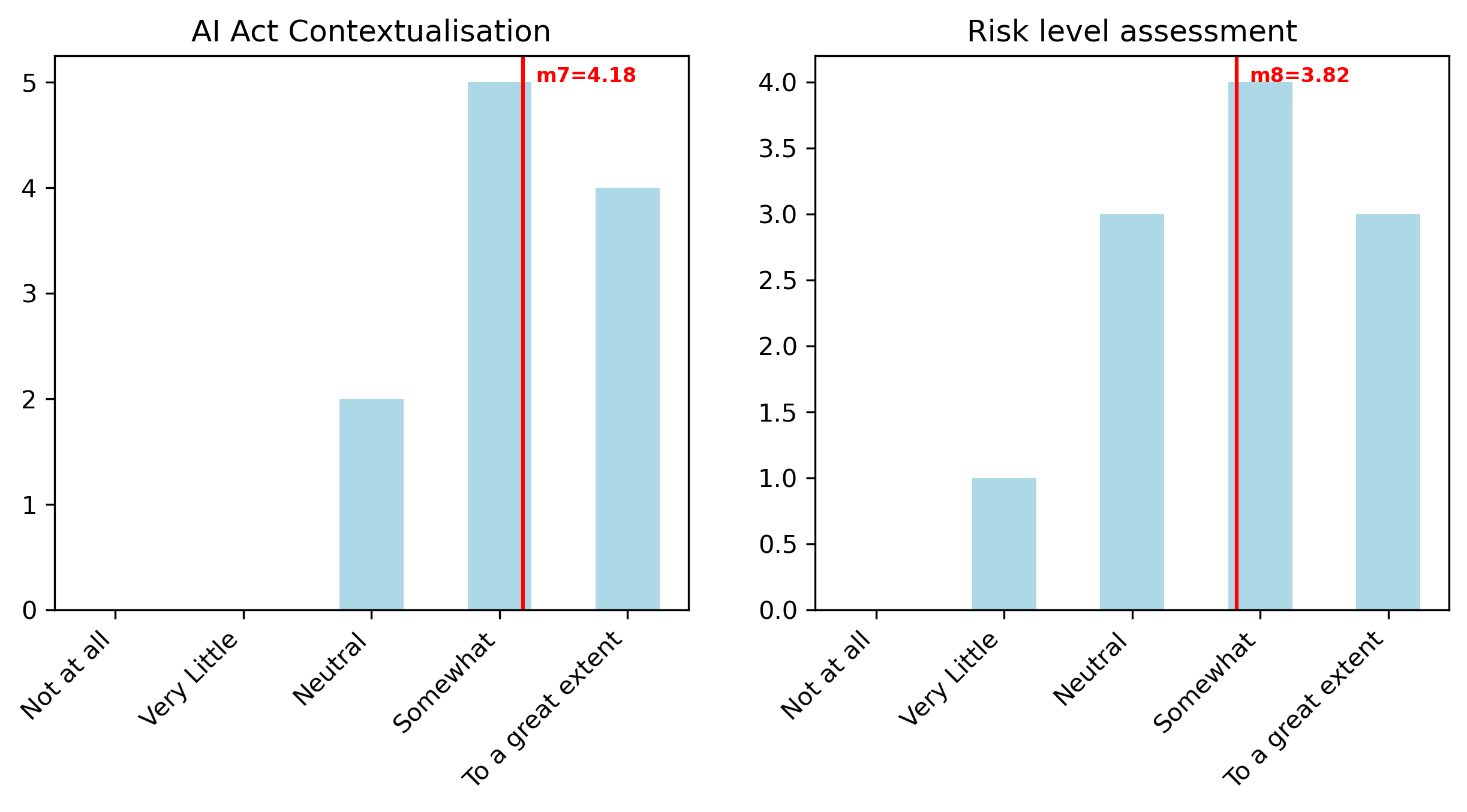}
  \caption{Histograms of the answers to questions 7 and 8, and mean values.}
  \label{fig:q7q8}
\end{figure}

\begin{figure}
\centering
  \includegraphics[width=0.9\textwidth]{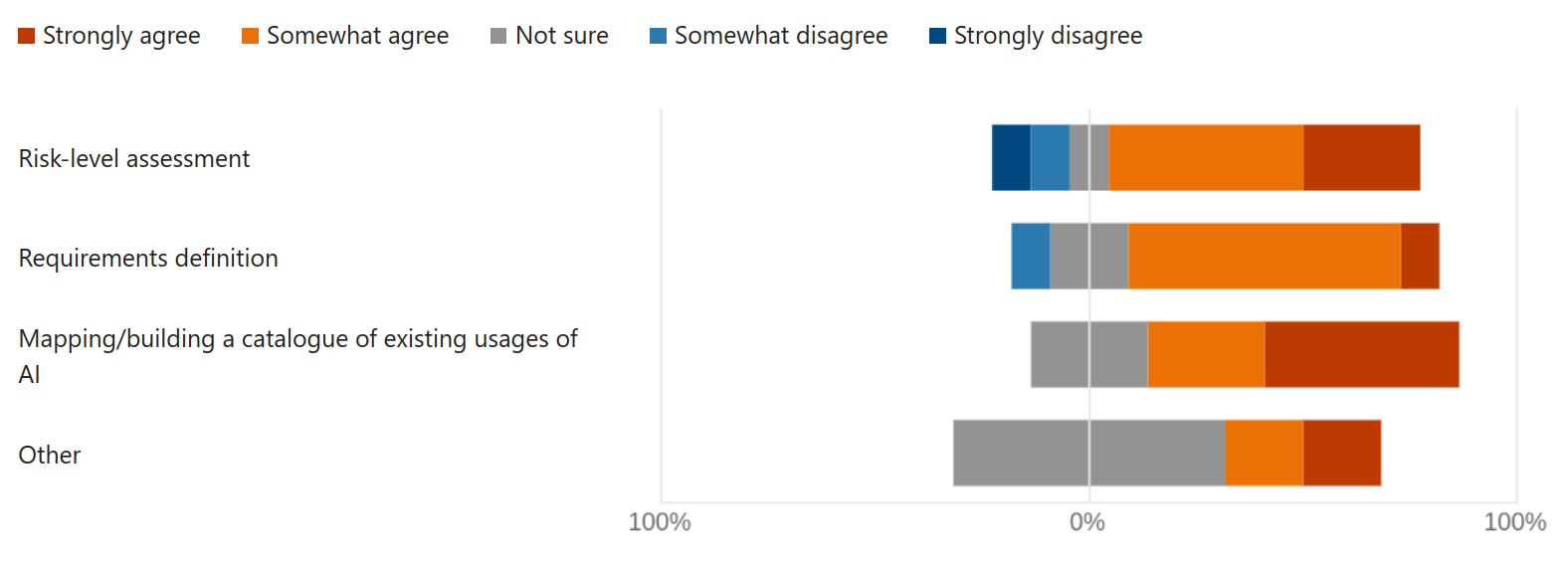}
  \caption{Visualization of answers to question 9 (``\textit{In the context of the AI Act, use case card is appropriate for...}'').}
  \label{fig:q9}
\end{figure}

Concerning the alignment of \textit{use case cards} with the AI Act, the feedback from the participants is also very positive. For example, regarding the level of contextualisation with the AI Act (question 7, Figure~\ref{fig:q7q8} left), the mean answer is between “somewhat” and “to a great extent”, with $M7=4.18$. Regarding its utility to assess the risk-level (question 8,  Figure \ref{fig:q7q8} right) the answers are between “very little” and “to a great extend”, with a mean value very close to “somewhat” ($M8=3.82$). And the general feedback from question 9 (Figure \ref{fig:q9}) is mostly positive towards an agreement on its appropriateness to different AI Act specific aspects. 

From the participants' answers to open question 3, we highlight the following suggestions for other potential uses:

\begin{itemize}
\item{“\textit{Documentation and training}”.}
\item{“\textit{As a standard to show the use of AI systems to citizens}”.}
\item{“\textit{Compare similar AI systems}”.}
\item{“\textit{Create a database of sample use cases}”.}
\item{“\textit{For conformity assessment}”.}
\item{“\textit{Elaborating on possible mitigation measures after risk assessment}”.}
\item{“\textit{To help non-experts to understand how a product works}”.}
\end{itemize}

Some of these answers echo our goal of proposing a methodology for documenting use cases for AI systems that is easy to understand by a non-expert audience. Other answers also point in the direction of a possible standard that could help with documentation needs, risk mitigation or conformity assessment.  

However, there are also some issues raised by some participants in the last open question. In almost all cases, the feedback obtained refers in one way or another to a limited expertise on 
UML for documenting use cases. For example, some participants did not clearly understood the difference between the “AI system” and the “use cases”, including some confusion about the type of dependencies between the use cases. This issue is highly correlated with the lack of previous knowledge on UML. Difficulties in learning and using UML are well-known issues in the research and industry communities~\citep{Siau2006}. However, the benefits of UML have been empirically validated in multiple studies~\citep{Chaudron2012}. While we recognise the potential initial difficulties of a wider audience in interpreting the UML canvas, we do not expect a major impact for AI providers, as UML is a \textit{de facto} industry standard for modelling software systems. Moreover, as most of the participants emphasised, the table is the main element of the proposed approach, and its clarity has been validated regardless of prior knowledge about UML.

\section{Conclusions}
\label{sec:conclusions}

In this work we present \textit{use case cards}, a standardised methodology for the documentation of AI use cases. It is grounded on four strong pillars: (1) the UML use case modelling standard; (2) the recently proposed European AI Act; (3) the result of a co-design with high-profile stakeholders including European 
 policy and scientific experts with a proficiency level on AI, UML and the AI Act; and (4) a validation with 11 experts combining technical knowledge on AI, social sciences, human rights and/or legal background, and having a strong experience in 
 EU digital policies.

Differently from other widely used methodologies for AI documentation, such as \textit{Model Cards}~\citep{mitchell2019model}, \textit{Method Cards}~\citep{adkins2022method} or \textit{System cards}~\citep{wahle2023ai}, \textit{use case cards} focuses on describing the intended purpose and operational use of an AI system rather than on the technical aspects related to --in most cases, a generic-- AI model. This allows to frame and put the use case in context, in a highly visual, complete and efficient manner. It has also be proven a useful tool for both policy makers and providers in assessing the risk level of an AI system, which is key to determine the legal obligations to which it must be subject.

It is important to emphasize nevertheless that \textit{use case cards} is not meant to be a final and exhaustive documentation methodology for compliance with any future legal requirement. First, because the AI Act is still under negotiation and therefore subject to possible modifications in its road towards adoption. Second, because the objective of this work is the documentation of use cases, which is just a small piece of the technical documentation required to demonstrate full conformity with the legal text.

\textit{Use case cards} has the potential to serve as a standardised methodology for documenting  
for use cases in the context of the European AI Act, as stated by participants in the co-design and validation exercises.  
In the future, we plan to develop a web-based prototype of this registry integrating a machine-editable version of \textit{use case cards} and allowing for the automated analysis of related statistics such as the number of use cases per application area, per product type, and most covered SDGs.

\backmatter



\section*{Declarations} 



\bmhead{Ethics approval}
The methodology followed in the user study was subject to ethical and data protection reviews in the context of the HUMAINT project, Joint Research Centre.  All participants both in the co-design phase and the questionnaire-based study were given an informed consent form with details regarding the purpose of the study, procedures and confidentiality treatment issues. They all voluntarily agreed to participate in the study.


\bmhead{Availability of data and materials} 
All data and material used and analysed in this study are available upon reasonable request to the corresponding author. Additionally the \textit{use case card} template and related examples are available at the public GitLab repository {\url{https://gitlab.com/humaint-ec_public/use-case-cards}}.




\newpage 

\begin{appendices}

\section{Lists of SDGs, products and application areas}\label{annex_A}

\begin{figure}[h!]
\centering
  \includegraphics[width=0.75\textwidth]{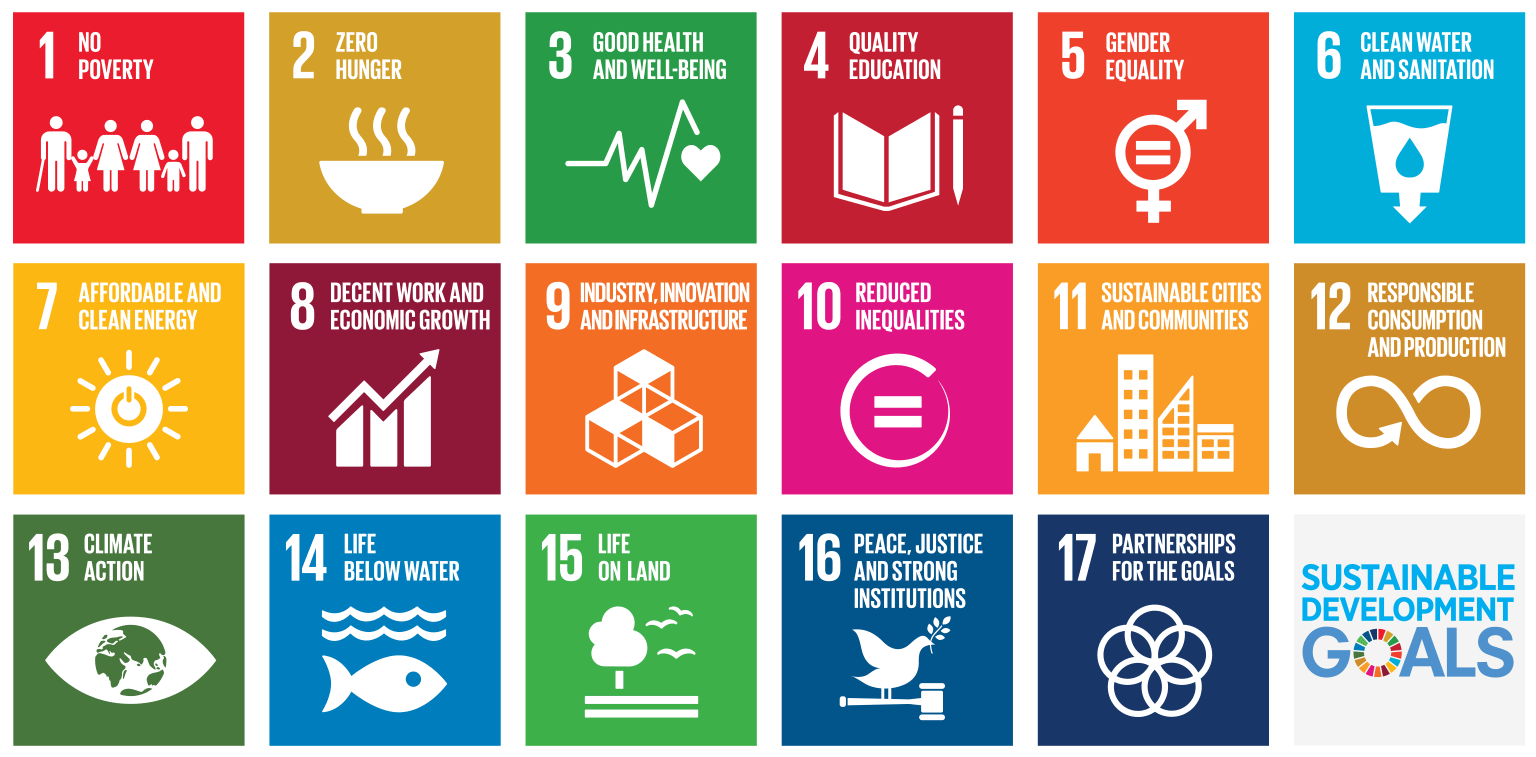}
  \caption{List of 17 Sustainable Development Goals (SDGs) defined by the United Nations~\citep{SDG}.}
  \label{fig:SDGs}
\end{figure}

This appendix lists the Sustainable Development Goals (SDGs, Figure~\ref{fig:SDGs}), type of products (Table~\ref{tab:products}) and application areas (Table~\ref{tab:areas}) to be used to fill in \textit{use case cards} as in Section~\ref{sec:ucc}.

\newcolumntype{W}{>{\raggedright\arraybackslash}m{1.16\textwidth}}
\newcolumntype{E}{>{\raggedright\arraybackslash}m{0.005\textwidth}}
\begin{table}[h!]
\centering
\footnotesize
\begin{tabular}{El}
\toprule
\multicolumn{2}{l}{\textbf{Type of product}}                                    \\
\toprule
\CircleHigh & Machinery                                                                       \\
\CircleHigh & Toy                                                                             \\
\CircleHigh & Recreational   craft or personal watercraft                                     \\
\CircleHigh & Lift                                                                            \\
\CircleHigh & Equipment and protective   systems for use in potentially explosive atmospheres \\
\CircleHigh & Radio equipment                                                                 \\
\CircleHigh & Pressure   equipment                                                            \\
\CircleHigh & Cableway   installation                                                         \\
\CircleHigh & Personal   protective equipment                                                 \\
\CircleHigh & Appliances   burning gaseous fuels                                              \\
\CircleHigh & Medical device                                                                  \\
\CircleHigh & In vitro diagnostic medical device                                            \\
\CircleHigh & Civil aviation                                                                  \\
\CircleHigh & 2- or 3-wheel   vehicle or quadricycle                                          \\
\CircleHigh & Agricultural and forestry vehicle                                             \\
\CircleHigh & Marine equipment                                                              \\
\CircleHigh & Interoperability of the rail system                                           \\
\CircleHigh & Motor vehicles and their trailers                                             \\
 & Other hardware  product/system                                                 \\
 & Other software  product/system                                                 \\ 
\bottomrule                                       
\end{tabular}
\caption{List of possible types of products for \textit{use case cards}. Those marked with \CircleHigh \hspace{0.2mm} might be subject to other European Union harmonisation legislation and, as such, be considered high-risk according to AI Act's Annex II (as of AI Act's ``General Approach'', December 2022).}
\label{tab:products}
\end{table}

\begin{table}[h!]
\centering
\footnotesize
\scalebox{0.8}{
\begin{tabular}{EW}
\toprule
\multicolumn{2}{l}{\textbf{Type of application area}}   \\
\toprule
\multicolumn{2}{l}{\textbf{Biometrics}} \\
\CircleHigh & Remote biometric identification systems.  \\
\midrule
\multicolumn{2}{l}{\textbf{Critical infrastructure}}  \\
\CircleHigh & AI systems used as safety components in the management and operation of critical digital infrastructure, road traffic and the supply of water, gas, heating and electricity.    \\
\midrule
\multicolumn{2}{l}{\textbf{Education and vocational training}}   \\
\CircleHigh & AI systems used to determine access, admission or to assign natural persons to educational and vocational training institutions or programmes.   \\
\cmidrule{2-2}
\CircleHigh  &  AI systems intended to be used to evaluate learning outcomes.  \\
\midrule
\multicolumn{2}{l}{\textbf{Employment, workers management and access to self-employment}}  \\
\CircleHigh  & AI systems used for recruitment or selection of natural persons, notably to place targeted job advertisements, to analyse and filter job applications, and to evaluate candidates.  \\
\cmidrule{2-2}
\CircleHigh &  AI systems to make decisions on promotion and termination of work-related relationships, to allocate tasks or monitor and evaluate performance based on person's behavior, personal traits or characteristics.    \\
\midrule
\multicolumn{2}{l}{\textbf{Access to essential private services, public services and benefits}}  \\
\CircleHigh &  AI systems used by public authorities to evaluate the eligibility of natural persons for essential public assistance benefits and services, and to grant, reduce, revoke or reclaim such benefits and services.  \\
\cmidrule{2-2}
\CircleHigh &  AI systems used to evaluate the creditworthiness of natural persons or establish their credit score.    \\
\cmidrule{2-2}
\CircleHigh &  AI systems used to dispatch, or to establish priority in the dispatching of emergency first response services, including by firefighters and medical aid.  \\
\cmidrule{2-2}
\CircleHigh &  AI systems for risk assessment and pricing in the case of life and health insurance. \\
\midrule
\multicolumn{2}{l}{\textbf{Law enforcement}}   \\
\CircleHigh &  AI systems  used by law enforcement to assess the risk of a natural person for offending or reoffending or the risk for a natural person to become a potential victim of criminal offences.  \\
\cmidrule{2-2}
\CircleHigh  &  AI systems used by law enforcement as polygraphs or to detect the emotional state of a natural person.    \\
\cmidrule{2-2}
\CircleHigh &  AI systems used by law enforcement to evaluate the reliability of evidence in the course of investigation or prosecution of criminal offences.    \\
\cmidrule{2-2}
\CircleHigh &  AI systems used by law enforcement to predict the (re)occurrence of a criminal offence based on profiling of natural persons or to assess personality traits and characteristics or past criminal behaviour.   \\
\cmidrule{2-2}
\CircleHigh  & AI systems used by law enforcement to profile natural persons in the course of detection, investigation or prosecution of criminal offences.\\
\midrule
\multicolumn{2}{l}{\textbf{Migration, asylum and border control management}}   \\
\CircleHigh  & AI systems used by public authorities  as polygraphs or to detect the emotional state of a natural  person.       \\
\cmidrule{2-2}
\CircleHigh  & AI systems used by public authorities to assess a risk (security risk, risk of irregular immigration, health risk) posed by a person who enters or has entered into the territory of a Member State.     \\
\cmidrule{2-2}
\CircleHigh  & AI systems to assist public authorities to examine applications for asylum, visa and residence permits and associated complaints.  \\
\midrule
\multicolumn{2}{l}{\textbf{Administration of justice and democratic processes}}  \\
\CircleHigh  & AI systems used by a judicial authority to interpret facts or the law and to apply the law to a concrete set of facts.  \\
\midrule
\multicolumn{2}{l}{\textbf{Entertainment and leisure}}   \\
\midrule
\multicolumn{2}{l}{\textbf{Marketing and retail}}   \\
\midrule
\multicolumn{2}{l}{\textbf{Culture, art and heritage}}    \\
\midrule
\multicolumn{2}{l}{\textbf{Clinical use in medicine and healthcare}}  \\
\midrule
\multicolumn{2}{l}{\textbf{Finances and banking}}  \\
\midrule
\multicolumn{2}{l}{\textbf{Social assistance}}  \\
\midrule
\multicolumn{2}{l}{\textbf{Video-surveillance for security}}  \\
\midrule
\multicolumn{2}{l}{\textbf{Transportation and mobility}}  \\
\midrule
\multicolumn{2}{l}{\textbf{Tourism, hospitality and restaurants}}    \\
\midrule
\multicolumn{2}{l}{\textbf{Industry and logistics} }   \\
\midrule
\multicolumn{2}{l}{\textbf{Politics}} \\
\midrule
\multicolumn{2}{l}{\textbf{Other}}   \\                                                                 \bottomrule                                                               
\end{tabular}
}
\caption{List of application areas for \textit{use case cards}. Subareas marked with \CircleHigh \hspace{0.2mm} are high-risk under AI Act's Annex III (as of AI Act's ``General Approach'', December 2022).}
\label{tab:areas}
\end{table}

\clearpage

\section{\textit{Use case cards} examples}\label{annex_B}

\begin{figure}[h!]
\centering
  \includegraphics[width=\textwidth]{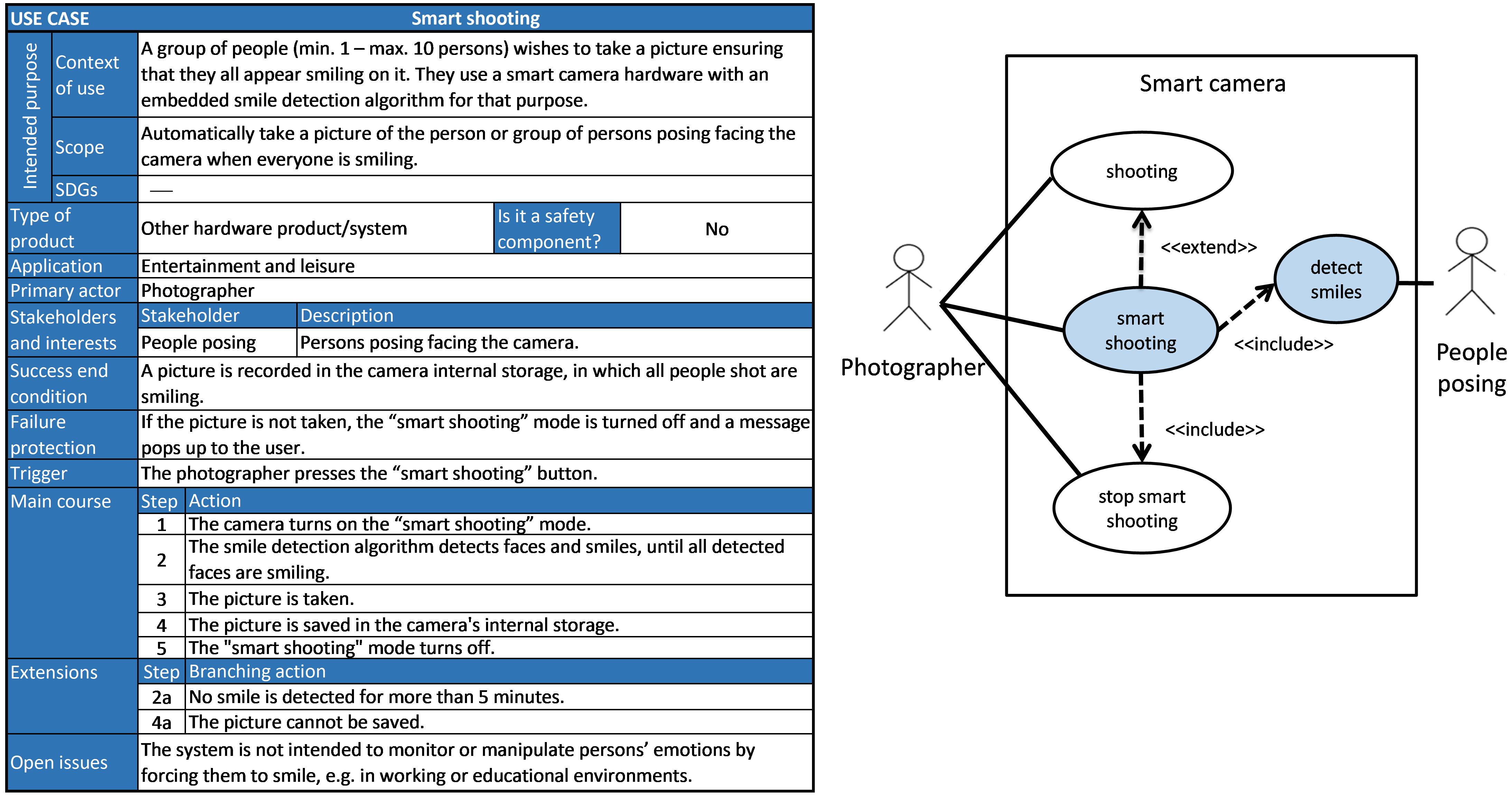}
  \caption{\textit{Use case card} for a smart camera system.}
  \label{fig:camera_ucc}
\end{figure}

\begin{figure}[h!]
\centering
  \includegraphics[width=\textwidth]{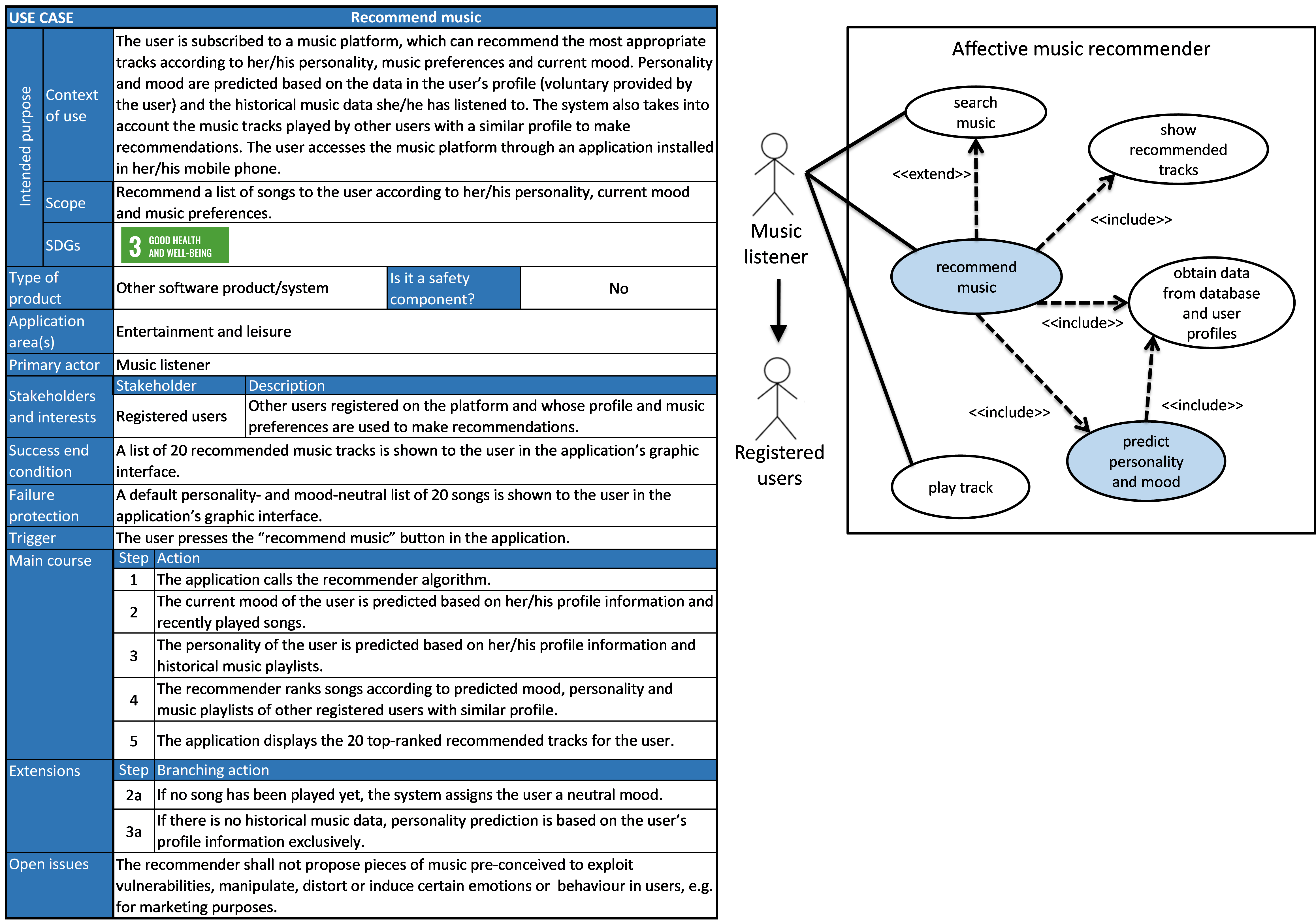}
  \caption{\textit{Use case card} for an affective music recommender system.}
  \label{fig:music_ucc}
\end{figure}

\begin{figure}[h!]
\centering
  \includegraphics[width=\textwidth]{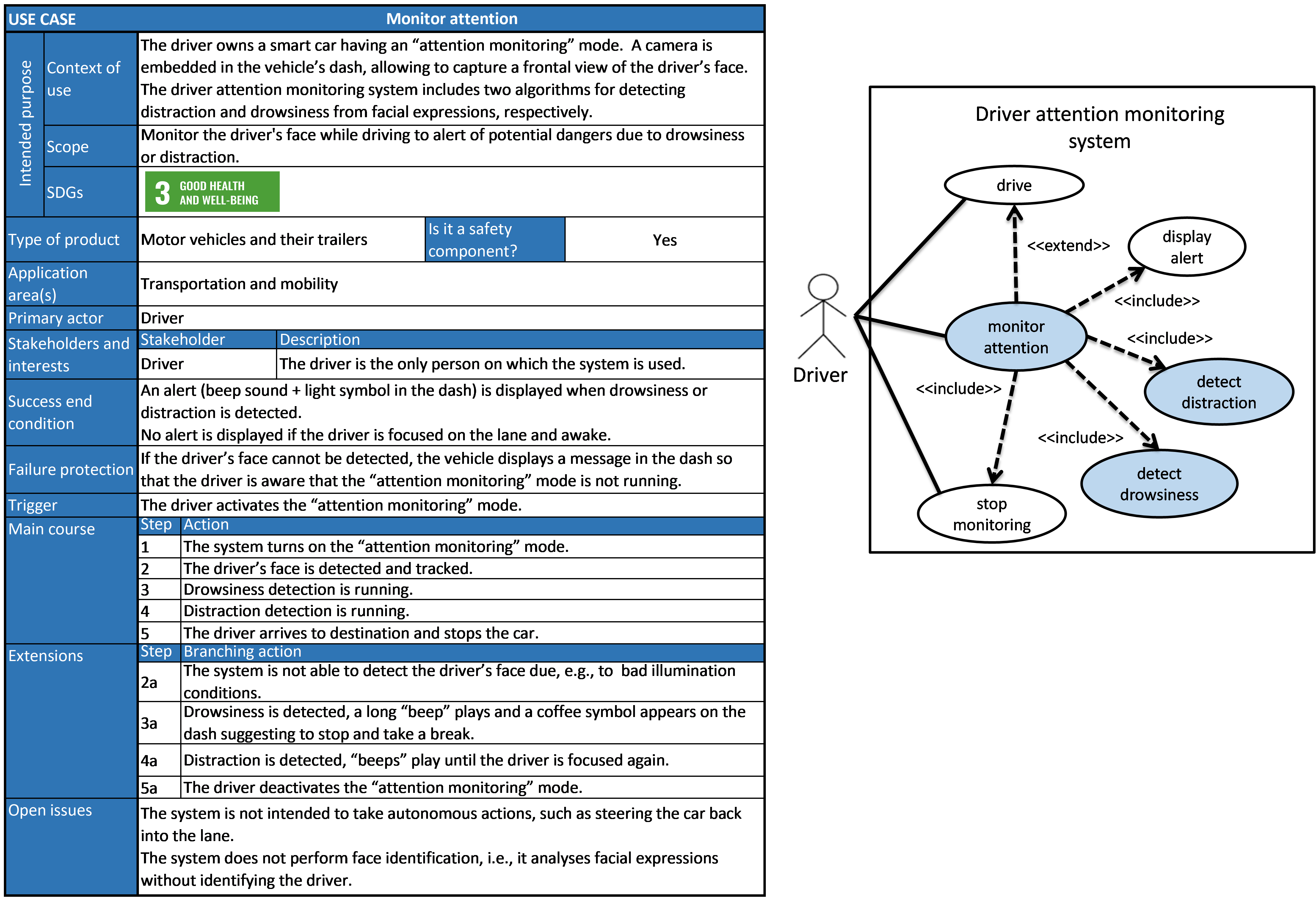}
  \caption{\textit{Use case card} for a driver attention monitoring system.}
  \label{fig:driver_monitoring_ucc}
\end{figure}

\begin{figure}[h!]
\centering
  \includegraphics[width=\textwidth]{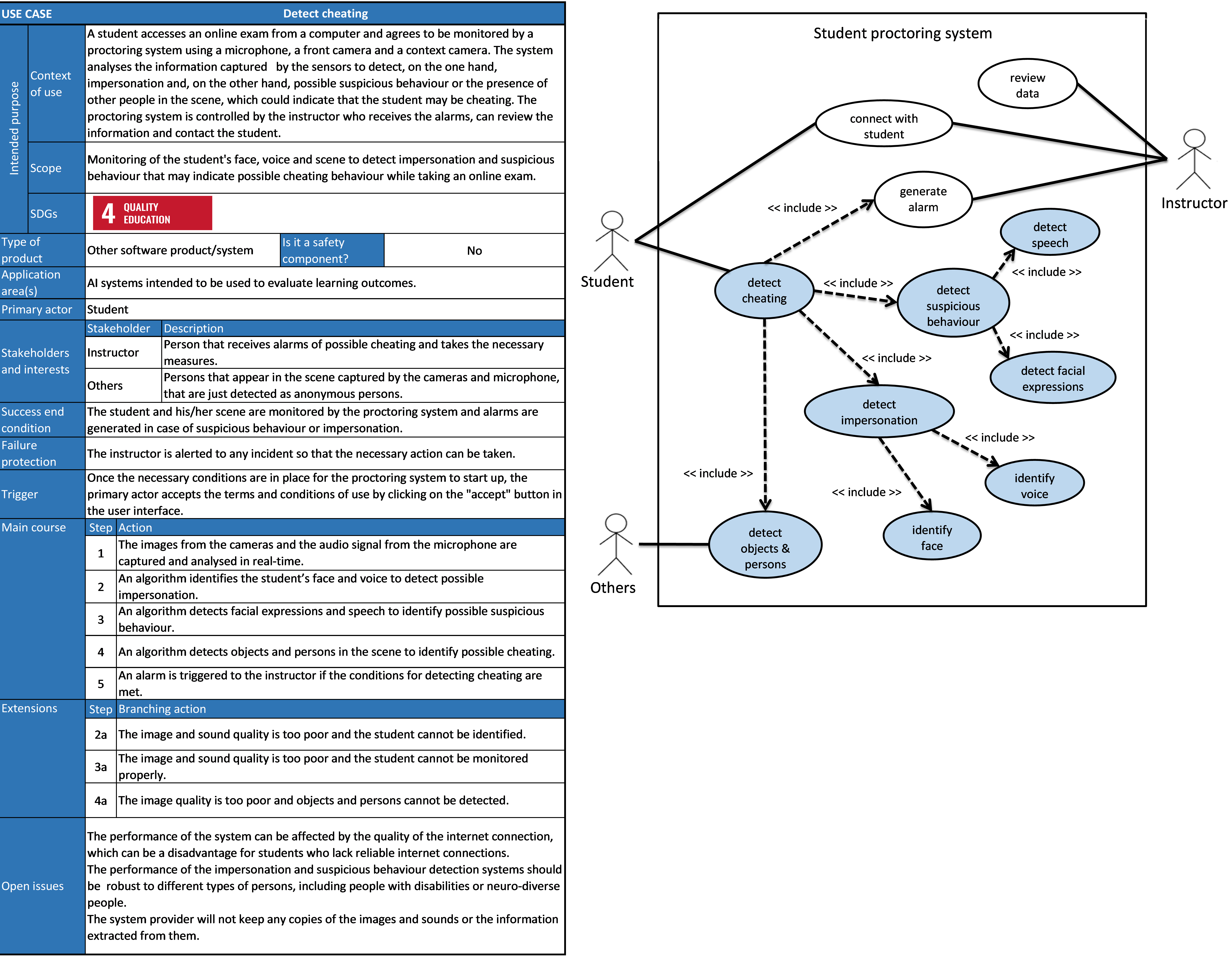}
  \caption{\textit{Use case card} for a student proctoring system.}
  \label{fig:proctoring_ucc}
\end{figure}

This annex presents four extra \textit{use case cards} examples. They were all developed with stakeholders during the co-design phase (Figures~\ref{fig:camera_ucc} to~\ref{fig:proctoring_ucc}). Two of them were additionally used in the questionnaire-based study (Figures~\ref{fig:driver_monitoring_ucc} and~\ref{fig:proctoring_ucc}). \\

\vspace{1cm}

\noindent \textbf{Smart camera}. In this example the AI-based system is a smart camera that shoots a picture only when all the people posing in front of it are smiling. There are several products in the market with this feature that serve as inspiration~\citep{canon,nikon}. The \textit{use case card} of the \textit{smart shooting} use case is shown in Figure~\ref{fig:camera_ucc}. This application is in principle simple and low-risk profile. However, it might lead to potential misuses that deserve documentation. For instance, a similar system was recently deployed in a working environment so that workers could only enter the front door or print documents when smiling to a camera. The management argued that it was intended to foster a positive working environment, but some workers felt their emotions were being manipulated~\citep{news_happyface}. \\

\newpage

\vspace*{0.5cm}

\noindent \textbf{Affective music recommender}. Figure~\ref{fig:music_ucc} shows the \textit{use case card} of a music recommender system proposing songs to the user based on personality, mood and playlist history. This use case has been inspired by~\cite{amini2019affective}. Several studies have demonstrated that music playlists can be used to infer user's emotions, personality traits and vulnerabilities~\citep{deshmukh2018survey}; the other way round, certain music pieces can induce behaviours and manipulate listeners' emotions~\citep{gomez2021music}. The \textit{use case card} allows to frame the ethical use of the system by stating that the sole purpose is providing the most appropriate music recommendations, and in any case manipulate listener's emotions or behaviour. \\

\vspace{0.5cm}

\newpage

\vspace*{0.7cm}

\noindent \textbf{Driver attention monitoring}. This AI system records a driver's face from a car's in-cabin camera and monitors facial behaviour to detect potential drowsiness and distraction. The \textit{monitor attention} use case is the one in charge of detecting such situations and sending alerts in the form of beep tones and light symbols in the car dash (Figure~\ref{fig:driver_monitoring_ucc}). Driver attention monitoring systems are nowadays commonly available as market products~\citep{subaru,tesla}. The corresponding \textit{use case card} states that the system is part of a safety component of the vehicle, which positions it as a high-risk system. Further, it highlights that the system is conceived to alert the driver but in any case to allow the vehicle to take full control of the car in an autonomous manner. \\

\vspace{0.7cm}

\newpage

\vspace*{0.7cm}

\noindent \textbf{Student proctoring}. This AI system detects potential cheating in students during exams. It is inspired by the literature~\citep{baldassarri2015affective,roa2022automated} and market products~\citep{procto_1,procto_2}. The \textit{use case card} presented in Figure~\ref{fig:proctoring_ucc} documents its main use case \textit{detect cheating}. It is a complex one as it includes AI computational tasks of different nature: video analysis for the detection of third persons in the room and relevant objects (e.g. books, phones); detection of impersonation through voice and face identification; and detection of suspicious behaviours (e.g. talking, facial/gaze movements). Alerts are triggered to instructors for review and action. This system's application area is high-risk and, as such, open issues such as ensuring non-discriminatory access and appropriate data governance must be carefully documented.

\vspace{0.7cm}

\end{appendices}


\clearpage

\bibliography{bibliography.bib}

\begin{thebibliography}{}
\providecommand{\doi}[1]{\url{https://doi.org/#1}}
\bibcommenthead

\bibitem [\protect \citeauthoryear {%
Adkins%
\ \protect \BOthers {.}}{%
Adkins%
\ \protect \BOthers {.}}{%
{\protect \APACyear {2022}}%
{\protect \APACexlab {{\protect \BCnt {1}}}}}]{%
adkins2022method}
\APACinsertmetastar {%
adkins2022method}%
\begin{APACrefauthors}%
Adkins, D.%
, Alsallakh, B.%
, Cheema, A.%
, Kokhlikyan, N.%
, McReynolds, E.%
, Mishra, P.%
\BDBL {}Zvyagina, P.%
\end{APACrefauthors}%
\unskip\
\newblock
\APACrefYearMonthDay{2022{\protect \BCnt {1}}}{}{}.
\newblock
{\BBOQ}\APACrefatitle {Method cards for prescriptive machine-learning
  transparency} {Method cards for prescriptive machine-learning
  transparency}.{\BBCQ}
\newblock
 \APACrefbtitle {{IEEE/ACM 1st International Conference on AI
  Engineering--Software Engineering for AI (CAIN)}} {{IEEE/ACM 1st
  International Conference on AI Engineering--Software Engineering for AI
  (CAIN)}}\ (\BPGS\ 90--100).
\PrintBackRefs{\CurrentBib}

\bibitem [\protect \citeauthoryear {%
Adkins%
\ \protect \BOthers {.}}{%
Adkins%
\ \protect \BOthers {.}}{%
{\protect \APACyear {2022}}%
{\protect \APACexlab {{\protect \BCnt {2}}}}}]{%
adkins2022prescriptive}
\APACinsertmetastar {%
adkins2022prescriptive}%
\begin{APACrefauthors}%
Adkins, D.%
, Alsallakh, B.%
, Cheema, A.%
, Kokhlikyan, N.%
, McReynolds, E.%
, Mishra, P.%
\BDBL {}Zvyagina, P.%
\end{APACrefauthors}%
\unskip\
\newblock
\APACrefYearMonthDay{2022{\protect \BCnt {2}}}{}{}.
\newblock
{\BBOQ}\APACrefatitle {Prescriptive and Descriptive Approaches to
  Machine-Learning Transparency} {Prescriptive and descriptive approaches to
  machine-learning transparency}.{\BBCQ}
\newblock
 \APACrefbtitle {{CHI} Conference on Human Factors in Computing Systems
  Extended Abstracts} {{CHI} conference on human factors in computing systems
  extended abstracts}\ (\BPGS\ 1--9).
\PrintBackRefs{\CurrentBib}

\bibitem [\protect \citeauthoryear {%
Amini%
, Willemsen%
\BCBL {}\ \BBA {} Graus%
}{%
Amini%
\ \protect \BOthers {.}}{%
{\protect \APACyear {2019}}%
}]{%
amini2019affective}
\APACinsertmetastar {%
amini2019affective}%
\begin{APACrefauthors}%
Amini, R.%
, Willemsen, M.C.%
\BCBL {} Graus, M.P.%
\end{APACrefauthors}%
\unskip\
\newblock
\APACrefYearMonthDay{2019}{}{}.
\newblock
{\BBOQ}\APACrefatitle {Affective Music Recommender System ({MRS}):
  Investigating the effectiveness and user satisfaction of different mood
  inducement strategies} {Affective music recommender system ({MRS}):
  Investigating the effectiveness and user satisfaction of different mood
  inducement strategies}.{\BBCQ}
\newblock

\newblock

\newblock

\PrintBackRefs{\CurrentBib}

\bibitem [\protect \citeauthoryear {%
Arnold%
\ \protect \BOthers {.}}{%
Arnold%
\ \protect \BOthers {.}}{%
{\protect \APACyear {2019}}%
}]{%
arnold2019factsheets}
\APACinsertmetastar {%
arnold2019factsheets}%
\begin{APACrefauthors}%
Arnold, M.%
, Bellamy, R.K.%
, Hind, M.%
, Houde, S.%
, Mehta, S.%
, Mojsilovi{\'c}, A.%
\BDBL {}others%
\end{APACrefauthors}%
\unskip\
\newblock
\APACrefYearMonthDay{2019}{}{}.
\newblock
{\BBOQ}\APACrefatitle {{AI FactSheets: Increasing trust in AI services through
  supplier's declarations of conformity}} {{AI FactSheets: Increasing trust in
  AI services through supplier's declarations of conformity}}.{\BBCQ}
\newblock
\APACjournalVolNumPages{IBM Journal of Research and
  Development}{63}{4/5}{6--1}.
\newblock

\newblock

\PrintBackRefs{\CurrentBib}

\bibitem [\protect \citeauthoryear {%
Baldassarri%
, Hupont%
, Abad{\'\i}a%
\BCBL {}\ \BBA {} Cerezo%
}{%
Baldassarri%
\ \protect \BOthers {.}}{%
{\protect \APACyear {2015}}%
}]{%
baldassarri2015affective}
\APACinsertmetastar {%
baldassarri2015affective}%
\begin{APACrefauthors}%
Baldassarri, S.%
, Hupont, I.%
, Abad{\'\i}a, D.%
\BCBL {} Cerezo, E.%
\end{APACrefauthors}%
\unskip\
\newblock
\APACrefYearMonthDay{2015}{}{}.
\newblock
{\BBOQ}\APACrefatitle {Affective-aware tutoring platform for interactive
  digital television} {Affective-aware tutoring platform for interactive
  digital television}.{\BBCQ}
\newblock
\APACjournalVolNumPages{Multimedia Tools and Applications}{74}{9}{3183--3206}.
\newblock

\newblock

\PrintBackRefs{\CurrentBib}

\bibitem [\protect \citeauthoryear {%
{Business Insider}%
}{%
{Business Insider}%
}{%
{\protect \APACyear {2021}}%
}]{%
news_happyface}
\APACinsertmetastar {%
news_happyface}%
\begin{APACrefauthors}%
{Business Insider}%
\end{APACrefauthors}%
\unskip\
\newblock
\APACrefYearMonthDay{2021}{}{}.
\newblock
\APACrefbtitle {{Employees at a Beijing office have to smile for an AI camera
  to get through the front doors, change the temperature, or print documents}.}
  {{Employees at a Beijing office have to smile for an AI camera to get through
  the front doors, change the temperature, or print documents}.}
\newblock
\APAChowpublished {Available:
  \url{https://www.businessinsider.com/workers-at-chinese-office-have-to-smile-at-ai-camera-2021-6}}.
\newblock
\APACrefnote{Online}
\PrintBackRefs{\CurrentBib}

\bibitem [\protect \citeauthoryear {%
{Canon}%
}{%
{Canon}%
}{%
{\protect \APACyear {2022}}%
}]{%
canon}
\APACinsertmetastar {%
canon}%
\begin{APACrefauthors}%
{Canon}%
\end{APACrefauthors}%
\unskip\
\newblock
\APACrefYearMonthDay{2022}{}{}.
\newblock
\APACrefbtitle {{Detecting a face and shooting (Smart Shutter)}.} {{Detecting a
  face and shooting (Smart Shutter)}.}
\newblock
\APAChowpublished {Available:
  \url{https://support.usa.canon.com/kb/index?page=content&id=ART133950}}.
\newblock
\APACrefnote{Online}
\PrintBackRefs{\CurrentBib}

\bibitem [\protect \citeauthoryear {%
Chaudron%
, Heijstek%
\BCBL {}\ \BBA {} Nugroho%
}{%
Chaudron%
\ \protect \BOthers {.}}{%
{\protect \APACyear {2012}}%
}]{%
Chaudron2012}
\APACinsertmetastar {%
Chaudron2012}%
\begin{APACrefauthors}%
Chaudron, M.%
, Heijstek, W.%
\BCBL {} Nugroho, A.%
\end{APACrefauthors}%
\unskip\
\newblock
\APACrefYearMonthDay{2012}{}{}.
\newblock
{\BBOQ}\APACrefatitle {How effective is UML modeling?} {How effective is uml
  modeling?}{\BBCQ}
\newblock
\APACjournalVolNumPages{Software \& Systems Modeling}{11}{}{571--580}.
\newblock

\newblock

\PrintBackRefs{\CurrentBib}

\bibitem [\protect \citeauthoryear {%
Chmielinski%
\ \protect \BOthers {.}}{%
Chmielinski%
\ \protect \BOthers {.}}{%
{\protect \APACyear {2022}}%
}]{%
chmielinski2022dataset}
\APACinsertmetastar {%
chmielinski2022dataset}%
\begin{APACrefauthors}%
Chmielinski, K.S.%
, Newman, S.%
, Taylor, M.%
, Joseph, J.%
, Thomas, K.%
, Yurkofsky, J.%
\BCBL {} Qiu, Y.C.%
\end{APACrefauthors}%
\unskip\
\newblock
\APACrefYearMonthDay{2022}{}{}.
\newblock
{\BBOQ}\APACrefatitle {The dataset nutrition label (2nd Gen): Leveraging
  context to mitigate harms in artificial intelligence} {The dataset nutrition
  label (2nd gen): Leveraging context to mitigate harms in artificial
  intelligence}.{\BBCQ}
\newblock
\APACjournalVolNumPages{arXiv preprint arXiv:2201.03954}{}{}{}.
\newblock

\newblock

\PrintBackRefs{\CurrentBib}

\bibitem [\protect \citeauthoryear {%
{Cloudsight}%
}{%
{Cloudsight}%
}{%
{\protect \APACyear {2023}}%
}]{%
narrator_2}
\APACinsertmetastar {%
narrator_2}%
\begin{APACrefauthors}%
{Cloudsight}%
\end{APACrefauthors}%
\unskip\
\newblock
\APACrefYearMonthDay{2023}{}{}.
\newblock
\APACrefbtitle {{TapTapSee App.}} {{TapTapSee App.}}
\newblock
\APAChowpublished {Available: \url{https://taptapseeapp.com/}}.
\newblock
\APACrefnote{Online}
\PrintBackRefs{\CurrentBib}

\bibitem [\protect \citeauthoryear {%
Cockburn%
}{%
Cockburn%
}{%
{\protect \APACyear {2001}}%
}]{%
cockburn2001writing}
\APACinsertmetastar {%
cockburn2001writing}%
\begin{APACrefauthors}%
Cockburn, A.%
\end{APACrefauthors}%
\unskip\
\newblock
\APACrefYear{2001}.
\newblock
\APACrefbtitle {Writing effective use cases} {Writing effective use cases}.
\newblock
\APACaddressPublisher{}{Pearson Education India}.
\PrintBackRefs{\CurrentBib}

\bibitem [\protect \citeauthoryear {%
{Council of the European Union}%
}{%
{Council of the European Union}%
}{%
{\protect \APACyear {2022}}%
}]{%
AIact_general}
\APACinsertmetastar {%
AIact_general}%
\begin{APACrefauthors}%
{Council of the European Union}%
\end{APACrefauthors}%
\unskip\
\newblock
\APACrefYearMonthDay{2022}{}{}.
\newblock
\APACrefbtitle {{Regulation on Artificial Intelligence - General approach}.}
  {{Regulation on Artificial Intelligence - General approach}.}
\newblock
\APAChowpublished {Available:
  \url{https://data.consilium.europa.eu/doc/document/ST-8115-2021-INIT/en/pdf}}.
\newblock
\APACrefnote{online}
\PrintBackRefs{\CurrentBib}

\bibitem [\protect \citeauthoryear {%
Deshmukh%
\ \BBA {} Kale%
}{%
Deshmukh%
\ \BBA {} Kale%
}{%
{\protect \APACyear {2018}}%
}]{%
deshmukh2018survey}
\APACinsertmetastar {%
deshmukh2018survey}%
\begin{APACrefauthors}%
Deshmukh, P.%
\BCBT {}\ \BBA {} Kale, G.%
\end{APACrefauthors}%
\unskip\
\newblock
\APACrefYearMonthDay{2018}{}{}.
\newblock
{\BBOQ}\APACrefatitle {A survey of music recommendation system} {A survey of
  music recommendation system}.{\BBCQ}
\newblock
\APACjournalVolNumPages{International Journal of Scientific Research in
  Computer Science, Engineering and Information Technology
  (IJSRCSEIT)}{3}{3}{1721--1729}.
\newblock

\newblock

\PrintBackRefs{\CurrentBib}

\bibitem [\protect \citeauthoryear {%
{European Commission}%
}{%
{European Commission}%
}{%
{\protect \APACyear {2019}}%
}]{%
HLEG}
\APACinsertmetastar {%
HLEG}%
\begin{APACrefauthors}%
{European Commission}%
\end{APACrefauthors}%
\unskip\
\newblock
\APACrefYearMonthDay{2019}{}{}.
\newblock
\APACrefbtitle {{Ethics Guidelines for Trustworthy AI}.} {{Ethics Guidelines
  for Trustworthy AI}.}
\newblock
\APAChowpublished {Available:
  \url{https://ec.europa.eu/digital-single-market/en/news/ethicsguidelines-trustworthy-ai}}.
\newblock
\APACrefnote{online}
\PrintBackRefs{\CurrentBib}

\bibitem [\protect \citeauthoryear {%
{European Commission}%
}{%
{European Commission}%
}{%
{\protect \APACyear {2021}}%
}]{%
AIact}
\APACinsertmetastar {%
AIact}%
\begin{APACrefauthors}%
{European Commission}%
\end{APACrefauthors}%
\unskip\
\newblock
\APACrefYearMonthDay{2021}{}{}.
\newblock
\APACrefbtitle {{Proposal for a Regulation of the European Parliament and of
  the Council laying down harmonised rules on Artificial Intelligence (AI Act)
  and amending certain union legislative acts}.} {{Proposal for a Regulation of
  the European Parliament and of the Council laying down harmonised rules on
  Artificial Intelligence (AI Act) and amending certain union legislative
  acts}.}
\newblock
\APAChowpublished {Available:
  \url{https://eur-lex.europa.eu/legal-content/EN/TXT/?uri=CELEX:52021PC0206}}.
\newblock
\APACrefnote{online}
\PrintBackRefs{\CurrentBib}

\bibitem [\protect \citeauthoryear {%
Fantechi%
, Gnesi%
, Lami%
\BCBL {}\ \BBA {} Maccari%
}{%
Fantechi%
\ \protect \BOthers {.}}{%
{\protect \APACyear {2003}}%
}]{%
fantechi2003applications}
\APACinsertmetastar {%
fantechi2003applications}%
\begin{APACrefauthors}%
Fantechi, A.%
, Gnesi, S.%
, Lami, G.%
\BCBL {} Maccari, A.%
\end{APACrefauthors}%
\unskip\
\newblock
\APACrefYearMonthDay{2003}{}{}.
\newblock
{\BBOQ}\APACrefatitle {Applications of linguistic techniques for use case
  analysis} {Applications of linguistic techniques for use case
  analysis}.{\BBCQ}
\newblock
\APACjournalVolNumPages{Requirements Engineering}{8}{3}{161--170}.
\newblock

\newblock

\PrintBackRefs{\CurrentBib}

\bibitem [\protect \citeauthoryear {%
Gebru%
\ \protect \BOthers {.}}{%
Gebru%
\ \protect \BOthers {.}}{%
{\protect \APACyear {2021}}%
}]{%
gebru2018datasheets}
\APACinsertmetastar {%
gebru2018datasheets}%
\begin{APACrefauthors}%
Gebru, T.%
, Morgenstern, J.%
, Vecchione, B.%
, Vaughan, J.W.%
, Wallach, H.%
, Iii, H.D.%
\BCBL {} Crawford, K.%
\end{APACrefauthors}%
\unskip\
\newblock
\APACrefYearMonthDay{2021}{}{}.
\newblock
{\BBOQ}\APACrefatitle {Datasheets for datasets} {Datasheets for
  datasets}.{\BBCQ}
\newblock
\APACjournalVolNumPages{Communications of the ACM}{64}{12}{86--92}.
\newblock

\newblock

\PrintBackRefs{\CurrentBib}

\bibitem [\protect \citeauthoryear {%
G{\'o}mez-Ca{\~n}{\'o}n%
\ \protect \BOthers {.}}{%
G{\'o}mez-Ca{\~n}{\'o}n%
\ \protect \BOthers {.}}{%
{\protect \APACyear {2021}}%
}]{%
gomez2021music}
\APACinsertmetastar {%
gomez2021music}%
\begin{APACrefauthors}%
G{\'o}mez-Ca{\~n}{\'o}n, J.S.%
, Cano, E.%
, Eerola, T.%
, Herrera, P.%
, Hu, X.%
, Yang, Y\BHBI H.%
\BCBL {} G{\'o}mez, E.%
\end{APACrefauthors}%
\unskip\
\newblock
\APACrefYearMonthDay{2021}{}{}.
\newblock
{\BBOQ}\APACrefatitle {Music emotion recognition: Toward new, robust standards
  in personalized and context-sensitive applications} {Music emotion
  recognition: Toward new, robust standards in personalized and
  context-sensitive applications}.{\BBCQ}
\newblock
\APACjournalVolNumPages{IEEE Signal Processing Magazine}{38}{6}{106--114}.
\newblock

\newblock

\PrintBackRefs{\CurrentBib}

\bibitem [\protect \citeauthoryear {%
{Google}%
}{%
{Google}%
}{%
{\protect \APACyear {2023}}%
}]{%
narrator_3}
\APACinsertmetastar {%
narrator_3}%
\begin{APACrefauthors}%
{Google}%
\end{APACrefauthors}%
\unskip\
\newblock
\APACrefYearMonthDay{2023}{}{}.
\newblock
\APACrefbtitle {{Lookout - Assisted vision application}.} {{Lookout - Assisted
  vision application}.}
\newblock
\APAChowpublished {Available:
  \url{https://play.google.com/store/apps/details?id=com.google.android.apps.accessibility.reveal&hl=en_US}}.
\newblock
\APACrefnote{Online}
\PrintBackRefs{\CurrentBib}

\bibitem [\protect \citeauthoryear {%
Gupta%
, Anpalagan%
, Guan%
\BCBL {}\ \BBA {} Khwaja%
}{%
Gupta%
\ \protect \BOthers {.}}{%
{\protect \APACyear {2021}}%
}]{%
gupta2021deep}
\APACinsertmetastar {%
gupta2021deep}%
\begin{APACrefauthors}%
Gupta, A.%
, Anpalagan, A.%
, Guan, L.%
\BCBL {} Khwaja, A.S.%
\end{APACrefauthors}%
\unskip\
\newblock
\APACrefYearMonthDay{2021}{}{}.
\newblock
{\BBOQ}\APACrefatitle {Deep learning for object detection and scene perception
  in self-driving cars: Survey, challenges, and open issues} {Deep learning for
  object detection and scene perception in self-driving cars: Survey,
  challenges, and open issues}.{\BBCQ}
\newblock
\APACjournalVolNumPages{Array}{10}{}{100057}.
\newblock

\newblock

\PrintBackRefs{\CurrentBib}

\bibitem [\protect \citeauthoryear {%
Hupont%
\ \BBA {} Gomez%
}{%
Hupont%
\ \BBA {} Gomez%
}{%
{\protect \APACyear {2022}}%
}]{%
hupont2022documenting2}
\APACinsertmetastar {%
hupont2022documenting2}%
\begin{APACrefauthors}%
Hupont, I.%
\BCBT {}\ \BBA {} Gomez, E.%
\end{APACrefauthors}%
\unskip\
\newblock
\APACrefYearMonthDay{2022}{}{}.
\newblock
{\BBOQ}\APACrefatitle {Documenting use cases in the affective computing domain
  using Unified Modeling Language} {Documenting use cases in the affective
  computing domain using unified modeling language}.{\BBCQ}
\newblock
\APACjournalVolNumPages{arXiv preprint arXiv:2209.09666}{}{}{}.
\newblock

\newblock

\PrintBackRefs{\CurrentBib}

\bibitem [\protect \citeauthoryear {%
Hupont%
, Micheli%
, Delipetrev%
, G{\'o}mez%
\BCBL {}\ \BBA {} Soler~Garrido%
}{%
Hupont%
, Micheli%
\BCBL {}\ \protect \BOthers {.}}{%
{\protect \APACyear {2022}}%
}]{%
hupont2022documenting}
\APACinsertmetastar {%
hupont2022documenting}%
\begin{APACrefauthors}%
Hupont, I.%
, Micheli, M.%
, Delipetrev, B.%
, G{\'o}mez, E.%
\BCBL {} Soler~Garrido, J.%
\end{APACrefauthors}%
\unskip\
\newblock
\APACrefYearMonthDay{2022}{}{}.
\newblock
{\BBOQ}\APACrefatitle {Documenting high-risk {AI}: a European regulatory
  perspective} {Documenting high-risk {AI}: a european regulatory
  perspective}.{\BBCQ}
\newblock
\APAChowpublished {Available:
  \url{https://www.techrxiv.org/articles/preprint/Documenting_high-risk_AI_an_European_regulatory_perspective/20291046}}.
\newblock
\APACrefnote{online}
\newblock

\newblock

\PrintBackRefs{\CurrentBib}

\bibitem [\protect \citeauthoryear {%
Hupont%
, Tolan%
, Gunes%
\BCBL {}\ \BBA {} G{\'o}mez%
}{%
Hupont%
, Tolan%
\BCBL {}\ \protect \BOthers {.}}{%
{\protect \APACyear {2022}}%
}]{%
hupont2022landscape}
\APACinsertmetastar {%
hupont2022landscape}%
\begin{APACrefauthors}%
Hupont, I.%
, Tolan, S.%
, Gunes, H.%
\BCBL {} G{\'o}mez, E.%
\end{APACrefauthors}%
\unskip\
\newblock
\APACrefYearMonthDay{2022}{}{}.
\newblock
{\BBOQ}\APACrefatitle {The Landscape of Facial Processing Applications in the
  Context of the European {AI} {A}ct and the Development of Trustworthy
  Systems} {The landscape of facial processing applications in the context of
  the european {AI} {A}ct and the development of trustworthy systems}.{\BBCQ}
\newblock
\APACjournalVolNumPages{Nature Scientific Reports}{}{}{}.
\newblock

\newblock

\PrintBackRefs{\CurrentBib}

\bibitem [\protect \citeauthoryear {%
Ko{\c{c}}%
, Erdo{\u{g}}an%
, Barjakly%
\BCBL {}\ \BBA {} Peker%
}{%
Ko{\c{c}}%
\ \protect \BOthers {.}}{%
{\protect \APACyear {2021}}%
}]{%
kocc2021uml}
\APACinsertmetastar {%
kocc2021uml}%
\begin{APACrefauthors}%
Ko{\c{c}}, H.%
, Erdo{\u{g}}an, A.M.%
, Barjakly, Y.%
\BCBL {} Peker, S.%
\end{APACrefauthors}%
\unskip\
\newblock
\APACrefYearMonthDay{2021}{}{}.
\newblock
{\BBOQ}\APACrefatitle {{UML} diagrams in software engineering research: a
  systematic literature review} {{UML} diagrams in software engineering
  research: a systematic literature review}.{\BBCQ}
\newblock
\APACjournalVolNumPages{Multidisciplinary Digital Publishing Institute
  Proceedings}{74}{1}{13}.
\newblock

\newblock

\PrintBackRefs{\CurrentBib}

\bibitem [\protect \citeauthoryear {%
Laato%
, Tiainen%
, Najmul~Islam%
\BCBL {}\ \BBA {} M{\"a}ntym{\"a}ki%
}{%
Laato%
\ \protect \BOthers {.}}{%
{\protect \APACyear {2022}}%
}]{%
laato2022explain}
\APACinsertmetastar {%
laato2022explain}%
\begin{APACrefauthors}%
Laato, S.%
, Tiainen, M.%
, Najmul~Islam, A.%
\BCBL {} M{\"a}ntym{\"a}ki, M.%
\end{APACrefauthors}%
\unskip\
\newblock
\APACrefYearMonthDay{2022}{}{}.
\newblock
{\BBOQ}\APACrefatitle {How to explain AI systems to end users: a systematic
  literature review and research agenda} {How to explain ai systems to end
  users: a systematic literature review and research agenda}.{\BBCQ}
\newblock
\APACjournalVolNumPages{Internet Research}{32}{7}{1--31}.
\newblock

\newblock

\PrintBackRefs{\CurrentBib}

\bibitem [\protect \citeauthoryear {%
Louradour%
\ \BBA {} Madzou%
}{%
Louradour%
\ \BBA {} Madzou%
}{%
{\protect \APACyear {2021}}%
}]{%
louradour2021policy}
\APACinsertmetastar {%
louradour2021policy}%
\begin{APACrefauthors}%
Louradour, S.%
\BCBT {}\ \BBA {} Madzou, L.%
\end{APACrefauthors}%
\unskip\
\newblock
\APACrefYearMonthDay{2021}{}{}.
\newblock
{\BBOQ}\APACrefatitle {A policy framework for responsible limits on facial
  recognition, Use case: Law enforcement investigations} {A policy framework
  for responsible limits on facial recognition, use case: Law enforcement
  investigations}.{\BBCQ}
\newblock
 \APACrefbtitle {World Economic Forum.} {World economic forum.}
\PrintBackRefs{\CurrentBib}

\bibitem [\protect \citeauthoryear {%
Madaio%
, Stark%
, Wortman~Vaughan%
\BCBL {}\ \BBA {} Wallach%
}{%
Madaio%
\ \protect \BOthers {.}}{%
{\protect \APACyear {2020}}%
}]{%
madaio2020co}
\APACinsertmetastar {%
madaio2020co}%
\begin{APACrefauthors}%
Madaio, M.A.%
, Stark, L.%
, Wortman~Vaughan, J.%
\BCBL {} Wallach, H.%
\end{APACrefauthors}%
\unskip\
\newblock
\APACrefYearMonthDay{2020}{}{}.
\newblock
{\BBOQ}\APACrefatitle {{Co-designing checklists to understand organizational
  challenges and opportunities around fairness in AI}} {{Co-designing
  checklists to understand organizational challenges and opportunities around
  fairness in AI}}.{\BBCQ}
\newblock
 \APACrefbtitle {{CHI} Conference on Human Factors in Computing Systems} {{CHI}
  conference on human factors in computing systems}\ (\BPGS\ 1--14).
\PrintBackRefs{\CurrentBib}

\bibitem [\protect \citeauthoryear {%
{Meazure Learning}%
}{%
{Meazure Learning}%
}{%
{\protect \APACyear {2023}}%
}]{%
procto_1}
\APACinsertmetastar {%
procto_1}%
\begin{APACrefauthors}%
{Meazure Learning}%
\end{APACrefauthors}%
\unskip\
\newblock
\APACrefYearMonthDay{2023}{}{}.
\newblock
\APACrefbtitle {{ProctorU proctoring platform.}} {{ProctorU proctoring
  platform.}}
\newblock
\APAChowpublished {Available: \url{https://www.proctoru.com}}.
\newblock
\APACrefnote{Online}
\PrintBackRefs{\CurrentBib}

\bibitem [\protect \citeauthoryear {%
{Meta}%
}{%
{Meta}%
}{%
{\protect \APACyear {2023}}%
}]{%
metaSC}
\APACinsertmetastar {%
metaSC}%
\begin{APACrefauthors}%
{Meta}%
\end{APACrefauthors}%
\unskip\
\newblock
\APACrefYearMonthDay{2023}{}{}.
\newblock
\APACrefbtitle {{System cards}.} {{System cards}.}
\newblock
\APAChowpublished {Available:
  \url{https://ai.facebook.com/tools/system-cards/}}.
\newblock
\APACrefnote{Online}
\PrintBackRefs{\CurrentBib}

\bibitem [\protect \citeauthoryear {%
{Microsoft}%
}{%
{Microsoft}%
}{%
{\protect \APACyear {2023}}%
}]{%
narrator_1}
\APACinsertmetastar {%
narrator_1}%
\begin{APACrefauthors}%
{Microsoft}%
\end{APACrefauthors}%
\unskip\
\newblock
\APACrefYearMonthDay{2023}{}{}.
\newblock
\APACrefbtitle {{Seeing AI App.}} {{Seeing AI App.}}
\newblock
\APAChowpublished {Available:
  \url{https://www.microsoft.com/en-us/ai/seeing-ai}}.
\newblock
\APACrefnote{Online}
\PrintBackRefs{\CurrentBib}

\bibitem [\protect \citeauthoryear {%
Mitchell%
\ \protect \BOthers {.}}{%
Mitchell%
\ \protect \BOthers {.}}{%
{\protect \APACyear {2019}}%
}]{%
mitchell2019model}
\APACinsertmetastar {%
mitchell2019model}%
\begin{APACrefauthors}%
Mitchell, M.%
, Wu, S.%
, Zaldivar, A.%
, Barnes, P.%
, Vasserman, L.%
, Hutchinson, B.%
\BDBL {}Gebru, T.%
\end{APACrefauthors}%
\unskip\
\newblock
\APACrefYearMonthDay{2019}{}{}.
\newblock
{\BBOQ}\APACrefatitle {Model cards for model reporting} {Model cards for model
  reporting}.{\BBCQ}
\newblock
 \APACrefbtitle {Conference on fairness, accountability, and transparency}
  {Conference on fairness, accountability, and transparency}\ (\BPGS\
  220--229).
\PrintBackRefs{\CurrentBib}

\bibitem [\protect \citeauthoryear {%
Nesbitt%
, Beleigoli%
, Du%
, Tirimacco%
\BCBL {}\ \BBA {} Clark%
}{%
Nesbitt%
\ \protect \BOthers {.}}{%
{\protect \APACyear {2022}}%
}]{%
nesbitt2022user}
\APACinsertmetastar {%
nesbitt2022user}%
\begin{APACrefauthors}%
Nesbitt, K.%
, Beleigoli, A.%
, Du, H.%
, Tirimacco, R.%
\BCBL {} Clark, R.A.%
\end{APACrefauthors}%
\unskip\
\newblock
\APACrefYearMonthDay{2022}{}{}.
\newblock
\APACrefbtitle {User experience (UX) Design as a co-design methodology: lessons
  learned during the development of a web-based portal for cardiac
  rehabilitation.} {User experience (ux) design as a co-design methodology:
  lessons learned during the development of a web-based portal for cardiac
  rehabilitation.}
\newblock
\APACaddressPublisher{}{Oxford University Press}.
\PrintBackRefs{\CurrentBib}

\bibitem [\protect \citeauthoryear {%
{Nikon}%
}{%
{Nikon}%
}{%
{\protect \APACyear {2022}}%
}]{%
nikon}
\APACinsertmetastar {%
nikon}%
\begin{APACrefauthors}%
{Nikon}%
\end{APACrefauthors}%
\unskip\
\newblock
\APACrefYearMonthDay{2022}{}{}.
\newblock
\APACrefbtitle {{Smile timer}.} {{Smile timer}.}
\newblock
\APAChowpublished {Available:
  \url{https://onlinemanual.nikonimglib.com/p950/en/#!/05-05}}.
\newblock
\APACrefnote{Online}
\PrintBackRefs{\CurrentBib}

\bibitem [\protect \citeauthoryear {%
{Object Management Group}%
}{%
{Object Management Group}%
}{%
{\protect \APACyear {2017}}%
}]{%
UML251}
\APACinsertmetastar {%
UML251}%
\begin{APACrefauthors}%
{Object Management Group}%
\end{APACrefauthors}%
\unskip\
\newblock
\APACrefYearMonthDay{2017}{}{}.
\newblock
\APACrefbtitle {{Unified Modeling Language specification v2.5.1}.} {{Unified
  Modeling Language specification v2.5.1}.}
\newblock
\APAChowpublished {Available: \url{https://www.omg.org/spec/UML/2.5.1/PDF}}.
\newblock
\APACrefnote{online}
\PrintBackRefs{\CurrentBib}

\bibitem [\protect \citeauthoryear {%
OECD%
}{%
OECD%
}{%
{\protect \APACyear {2022}}%
}]{%
OECD}
\APACinsertmetastar {%
OECD}%
\begin{APACrefauthors}%
OECD%
\end{APACrefauthors}%
\unskip\
\newblock
\APACrefYearMonthDay{2022}{}{}.
\newblock
\APACrefbtitle {{OECD Framework for Classification of AI Systems: a tool for
  effective AI policies}.} {{OECD Framework for Classification of AI Systems: a
  tool for effective AI policies}.}
\newblock
\APAChowpublished {Available: \url{https://oecd.ai/en/classification}}.
\newblock
\APACrefnote{online}
\PrintBackRefs{\CurrentBib}

\bibitem [\protect \citeauthoryear {%
Panigutti%
, Ronan%
\BCBL {}\ \BBA {} et al.%
}{%
Panigutti%
\ \protect \BOthers {.}}{%
{\protect \APACyear {2023}}%
}]{%
XAI_2023}
\APACinsertmetastar {%
XAI_2023}%
\begin{APACrefauthors}%
Panigutti, C.%
, Ronan, H.%
\BCBL {} et al.%
\end{APACrefauthors}%
\unskip\
\newblock
\APACrefYearMonthDay{2023}{}{}.
\newblock
{\BBOQ}\APACrefatitle {The role of explainable AI in the context of the AI Act}
  {The role of explainable ai in the context of the ai act}.{\BBCQ}
\newblock
 \APACrefbtitle {{6th ACM Conference on Fairness, Accountability and
  Transparency (FAccT)}.} {{6th ACM Conference on Fairness, Accountability and
  Transparency (FAccT)}.}
\PrintBackRefs{\CurrentBib}

\bibitem [\protect \citeauthoryear {%
Post%
}{%
Post%
}{%
{\protect \APACyear {2022}}%
}]{%
tesla}
\APACinsertmetastar {%
tesla}%
\begin{APACrefauthors}%
Post, T.W.%
\end{APACrefauthors}%
\unskip\
\newblock
\APACrefYearMonthDay{2022}{}{}.
\newblock
\APACrefbtitle {Full Self-Driving clips show owners of {T}eslas fighting for
  control, and experts see deep flaws.} {Full self-driving clips show owners of
  {T}eslas fighting for control, and experts see deep flaws.}
\newblock
\APAChowpublished
  {\url{https://www.washingtonpost.com/technology/2022/02/10/video-tesla-full-self-driving-beta/}}.
\newblock
\APACrefnote{online}
\PrintBackRefs{\CurrentBib}

\bibitem [\protect \citeauthoryear {%
Pushkarna%
, Zaldivar%
\BCBL {}\ \BBA {} Kjartansson%
}{%
Pushkarna%
\ \protect \BOthers {.}}{%
{\protect \APACyear {2022}}%
}]{%
pushkarna2022data}
\APACinsertmetastar {%
pushkarna2022data}%
\begin{APACrefauthors}%
Pushkarna, M.%
, Zaldivar, A.%
\BCBL {} Kjartansson, O.%
\end{APACrefauthors}%
\unskip\
\newblock
\APACrefYearMonthDay{2022}{}{}.
\newblock
{\BBOQ}\APACrefatitle {Data Cards: Purposeful and Transparent Dataset
  Documentation for Responsible AI} {Data cards: Purposeful and transparent
  dataset documentation for responsible ai}.{\BBCQ}
\newblock
\APACjournalVolNumPages{arXiv preprint arXiv:2204.01075}{}{}{}.
\newblock

\newblock

\PrintBackRefs{\CurrentBib}

\bibitem [\protect \citeauthoryear {%
{Respondus}%
}{%
{Respondus}%
}{%
{\protect \APACyear {2023}}%
}]{%
procto_2}
\APACinsertmetastar {%
procto_2}%
\begin{APACrefauthors}%
{Respondus}%
\end{APACrefauthors}%
\unskip\
\newblock
\APACrefYearMonthDay{2023}{}{}.
\newblock
\APACrefbtitle {{Assessment tools for learning systems.}} {{Assessment tools
  for learning systems.}}
\newblock
\APAChowpublished {Available: \url{https://web.respondus.com}}.
\newblock
\APACrefnote{Online}
\PrintBackRefs{\CurrentBib}

\bibitem [\protect \citeauthoryear {%
Roa’a%
, Aljazaery%
, ALRikabi%
\BCBL {}\ \BBA {} Alaidi%
}{%
Roa’a%
\ \protect \BOthers {.}}{%
{\protect \APACyear {2022}}%
}]{%
roa2022automated}
\APACinsertmetastar {%
roa2022automated}%
\begin{APACrefauthors}%
Roa’a, M.%
, Aljazaery, I.A.%
, ALRikabi, H.T.S.%
\BCBL {} Alaidi, A.H.M.%
\end{APACrefauthors}%
\unskip\
\newblock
\APACrefYearMonthDay{2022}{}{}.
\newblock
{\BBOQ}\APACrefatitle {Automated Cheating Detection based on Video Surveillance
  in the Examination Classes} {Automated cheating detection based on video
  surveillance in the examination classes}.{\BBCQ}
\newblock
\APACjournalVolNumPages{iJIM}{16}{08}{125}.
\newblock

\newblock

\PrintBackRefs{\CurrentBib}

\bibitem [\protect \citeauthoryear {%
S{\'a}nchez%
, Hupont%
, Tabik%
\BCBL {}\ \BBA {} Herrera%
}{%
S{\'a}nchez%
\ \protect \BOthers {.}}{%
{\protect \APACyear {2020}}%
}]{%
sanchez2020revisiting}
\APACinsertmetastar {%
sanchez2020revisiting}%
\begin{APACrefauthors}%
S{\'a}nchez, F.L.%
, Hupont, I.%
, Tabik, S.%
\BCBL {} Herrera, F.%
\end{APACrefauthors}%
\unskip\
\newblock
\APACrefYearMonthDay{2020}{}{}.
\newblock
{\BBOQ}\APACrefatitle {Revisiting crowd behaviour analysis through deep
  learning: Taxonomy, anomaly detection, crowd emotions, datasets,
  opportunities and prospects} {Revisiting crowd behaviour analysis through
  deep learning: Taxonomy, anomaly detection, crowd emotions, datasets,
  opportunities and prospects}.{\BBCQ}
\newblock
\APACjournalVolNumPages{Information Fusion}{64}{}{318--335}.
\newblock

\newblock

\PrintBackRefs{\CurrentBib}

\bibitem [\protect \citeauthoryear {%
Siau%
\ \BBA {} Loo%
}{%
Siau%
\ \BBA {} Loo%
}{%
{\protect \APACyear {2006}}%
}]{%
Siau2006}
\APACinsertmetastar {%
Siau2006}%
\begin{APACrefauthors}%
Siau, K.%
\BCBT {}\ \BBA {} Loo, P\BHBI P.%
\end{APACrefauthors}%
\unskip\
\newblock
\APACrefYearMonthDay{2006}{}{}.
\newblock
{\BBOQ}\APACrefatitle {{Identifying Difficulties in Learning Uml}}
  {{Identifying Difficulties in Learning Uml}}.{\BBCQ}
\newblock
\APACjournalVolNumPages{Information Systems Management}{23}{3}{43-51}.
\newblock
\begin{APACrefURL}
  {https://doi.org/10.1201/1078.10580530/46108.23.3.20060601/93706.5}
  \end{APACrefURL}
\newblock
{\href{https://arxiv.org/abs/https://doi.org/10.1201/1078.10580530/46108.23.3.20060601/93706.5}{{https://doi.org/10.1201/1078.10580530/46108.23.3.20060601/93706.5}}}
\newblock

\newblock
\begin{APACrefDOI} \doi{10.1201/1078.10580530/46108.23.3.20060601/93706.5}
  \end{APACrefDOI}
\PrintBackRefs{\CurrentBib}

\bibitem [\protect \citeauthoryear {%
{Subaru}%
}{%
{Subaru}%
}{%
{\protect \APACyear {2022}}%
}]{%
subaru}
\APACinsertmetastar {%
subaru}%
\begin{APACrefauthors}%
{Subaru}%
\end{APACrefauthors}%
\unskip\
\newblock
\APACrefYearMonthDay{2022}{}{}.
\newblock
\APACrefbtitle {{Driver monitoring system.}} {{Driver monitoring system.}}
\newblock
\APAChowpublished {Available:
  \url{https://www.subaru.com.au/driver-monitoring-system}}.
\newblock
\APACrefnote{Online}
\PrintBackRefs{\CurrentBib}

\bibitem [\protect \citeauthoryear {%
{United Nations}%
}{%
{United Nations}%
}{%
{\protect \APACyear {2023}}%
}]{%
SDG}
\APACinsertmetastar {%
SDG}%
\begin{APACrefauthors}%
{United Nations}%
\end{APACrefauthors}%
\unskip\
\newblock
\APACrefYearMonthDay{2023}{}{}.
\newblock
\APACrefbtitle {{Sustainable Development Goals}.} {{Sustainable Development
  Goals}.}
\newblock
\APAChowpublished {Available:
  \url{https://www.undp.org/sustainable-development-goals}}.
\newblock
\APACrefnote{Online}
\PrintBackRefs{\CurrentBib}

\bibitem [\protect \citeauthoryear {%
Wahle%
, Ruas%
, Mohammad%
, Meuschke%
\BCBL {}\ \BBA {} Gipp%
}{%
Wahle%
\ \protect \BOthers {.}}{%
{\protect \APACyear {2023}}%
}]{%
wahle2023ai}
\APACinsertmetastar {%
wahle2023ai}%
\begin{APACrefauthors}%
Wahle, J.P.%
, Ruas, T.%
, Mohammad, S.M.%
, Meuschke, N.%
\BCBL {} Gipp, B.%
\end{APACrefauthors}%
\unskip\
\newblock
\APACrefYearMonthDay{2023}{}{}.
\newblock
{\BBOQ}\APACrefatitle {{AI usage cards: Responsibly reporting AI-generated
  content}} {{AI usage cards: Responsibly reporting AI-generated
  content}}.{\BBCQ}
\newblock
\APACjournalVolNumPages{arXiv preprint arXiv:2303.03886}{}{}{}.
\newblock

\newblock

\PrintBackRefs{\CurrentBib}

\bibitem [\protect \citeauthoryear {%
Zamenopoulos%
\ \BBA {} Alexiou%
}{%
Zamenopoulos%
\ \BBA {} Alexiou%
}{%
{\protect \APACyear {2018}}%
}]{%
zamenopoulos2018co}
\APACinsertmetastar {%
zamenopoulos2018co}%
\begin{APACrefauthors}%
Zamenopoulos, T.%
\BCBT {}\ \BBA {} Alexiou, K.%
\end{APACrefauthors}%
\unskip\
\newblock
\APACrefYear{2018}.
\newblock
\APACrefbtitle {Co-design as collaborative research} {Co-design as
  collaborative research}.
\newblock
\APACaddressPublisher{}{Bristol University/AHRC Connected Communities
  Programme}.
\PrintBackRefs{\CurrentBib}

\end{thebibliography}

\end{document}